\documentclass{article}
\usepackage[utf8]{inputenc}
\usepackage{authblk}
\usepackage{setspace}
\usepackage[margin=1.25in]{geometry}
\usepackage{graphicx}
\graphicspath{{./}}
\usepackage{subcaption}
\usepackage{amsmath,amssymb,amsfonts}
\usepackage{lineno}
\usepackage{hyperref}
\usepackage{float}

\usepackage{array}
\usepackage{longtable}
\usepackage{tabularx}
\usepackage{microtype}

\usepackage{amsthm}
\newtheorem{definition}{Definition}

\usepackage[authoryear,round]{natbib}

\newcolumntype{P}[1]{>{\raggedright\arraybackslash}p{#1}}
\newcolumntype{Y}{>{\raggedright\arraybackslash}X}
\newcolumntype{S}[1]{>{\raggedright\arraybackslash\scriptsize}p{#1}} % compact sources col

\title{Range-Based Volatility Estimators for Monitoring Market Stress: Evidence from Local Food Price Data}
\author[1*$\dag$]{Bo Pieter Johannes Andr{\'e}e}

\affil[1]{Data Group, World Bank, Geneva, Switzerland.}

\date{January, 2026}

\begin{document}

\maketitle

\begin{abstract}
Range-based volatility estimators are widely used in financial econometrics to quantify risk and market stress, yet their application to local commodity markets remains limited. This paper shows how open--high--low--close (OHLC) volatility estimators can be adapted to monitor localized market distress across diverse development contexts, including conflict-affected settings, climate-exposed regions, remote and thinly traded markets, and import- and logistics-constrained urban hubs. Using monthly food price data from the World Bank's Real-Time Prices dataset, several volatility measures---including the Parkinson, Garman--Klass, Rogers--Satchell, and Yang--Zhang estimators---are constructed and evaluated against independently documented disruption timelines. Across settings, elevated volatility aligns with episodes linked to insecurity and market fragmentation, extreme weather and disaster shocks, policy and fuel-cost adjustments, and global supply-chain and trade disruptions. Volatility also detects stress that standard momentum indicators such as the relative strength index (RSI) can miss, including symmetric or rapidly reversing shocks in which offsetting supply and demand disturbances dampen net directional price movements while amplifying intra-period dispersion. Overall, OHLC-based volatility indicators provide a robust and interpretable signal of market disruptions and complement price-level monitoring for applications spanning financial risk, humanitarian early warning, and trade.\footnote{*$\dag$Bo Pieter Johannes Andrée, The World Bank, Development Economics, Data Group, can be contacted at bandree(at)worldbank.org. Funding by the World Bank's Food Systems 2030 (FS2030) Multi-Donor Trust Fund program (TF0C7822) and the Umbrella Facility for Trade Multi-Donor Trust Fund 2.0 (TF074184) is gratefully acknowledged. The findings, interpretations, and conclusions expressed in this paper are entirely those of the author. They do not necessarily represent the views of the World Bank and its affiliated organizations, or those of the Executive Directors of the World Bank or the governments they represent.}
\newline
\newline
\noindent \textbf{Keywords}: {Volatility estimation, Range-based estimators, Market stress indicators, Price monitoring, OHLC data. \vspace{1em}}
\end{abstract}

\pagebreak

\section{Introduction}

Volatility is a fundamental measure of market conditions. In financial markets, elevated volatility signals uncertainty and is frequently associated with reduced liquidity and impaired market functioning \citep{schwert1989,bollerslev1992,adrianetal2017}. A central insight from market microstructure is that well-functioning markets support continuous price discovery around an ``efficient price,'' while frictions such as higher transaction costs and widening spreads distort price adjustment and amplify short-horizon dispersion \citep{demsetz1968,hasbrouck1995,hasbrouck2002,roll1984,corwinschultz2012}. Financial econometrics offers a rich toolkit for measuring volatility from price data, including close-to-close estimators and more efficient range-based estimators that exploit information in intra-period highs and lows \citep{parkinson1980,garmanklass1980,rogerssatchell1991,yangzhang2000}. These measures are widely used in risk management, derivatives pricing, and trading strategies \citep{andersen2003,baillie1996,hasbroucksaar2013}.

Despite their established role in finance, OHLC-based range estimators have seen limited systematic application in monitoring local commodity markets in developing countries. This gap is consequential. In settings exposed to conflict, climate shocks, policy disruptions, and global price transmission, volatility can reveal stress that price levels alone may not capture. Market integration can weaken even when average prices move gradually, as disruptions raise arbitrage costs, reduce trading depth, and increase uncertainty about supply and demand. While the institutional settings differ from financial markets, the economic logic is analogous: in both contexts, elevated transaction costs and reduced market depth manifest as increased short-horizon price dispersion. Dysfunction is more likely to appear as widening spreads, abrupt reversals, and elevated volatility, echoing patterns documented for stressed financial markets \citep{copelandgalai1983,flemingremolona1999,baoetal2011}. This distinction matters for market monitoring: price levels primarily track changes in purchasing power, whereas volatility extracted from intra-period dispersion provides complementary information about frictions in market clearing and price discovery. In particular, compound shocks that simultaneously affect supply and demand can dampen net price changes while amplifying intra-period dispersion, making volatility an informative indicator of market stress.

The case for volatility monitoring is especially strong in food markets, where trading volumes and flows are unobserved and prices alone paint an incomplete picture. Food price instability is widely recognized as a macroeconomic and food-security concern \citep{fao2009,hlpe2011,gilbert2010}, and empirical work links volatility to food-security outcomes and risk transmission across markets \citep{benabdallah2021,lopezcabrera2013,hanif2021}. Yet many operational tools remain centered on price levels and trend-following momentum indicators such as RSI and MACD (moving average convergence divergence) \citep{appel1979,murphy1999,wilder1978}, which can understate short-lived reversals and symmetric shocks that manifest primarily through instability rather than persistent directional movement.

The need for robust, lightweight signals is growing as global monitoring systems increasingly rely on near-real-time indicators to support decision-making. Volatility-based surveillance has long been standard practice in financial markets, where exchanges, clearing houses, and regulators routinely monitor market functionality and liquidity. Macroeconomic monitoring systems and early warning systems for food security have been slower to adopt such measures, despite operating in environments where market dysfunction carries immediate welfare consequences \citep{barrett2010_measuring}. Current operational platforms integrate climate, production, market, and conflict information to guide timely assessments \citep{backer2021,bertetti2024}, and the World Bank's Real-Time Prices (RTP) initiative complements these efforts by harmonizing partner price observations into monthly OHLC series across dozens of countries \citep{rtp2021,andree2021_inflation,andreepape2023}---yet volatility extracted from these series remains underutilized.

This paper demonstrates that OHLC-based volatility estimators applied to local market price series provide a robust and interpretable signal of market distress across diverse settings. We compute close-to-close and range-based estimators directly from RTP's OHLC series, compare their operational properties, and apply a transparent threshold rule to flag stress episodes. Across the estimators considered, the Yang--Zhang measure \citep{yangzhang2000} offers a practical balance between responsiveness and noise reduction, and the detected episodes align closely with documented shocks. We illustrate the approach through five applications spanning distinct regions, market structures, and shock profiles: Sudan (Darfur), Somalia (Baidoa), Cameroon (Far North), Haiti (Port-au-Prince), and the Philippines (Sulu). Across these settings, volatility spikes coincide with conflict dynamics, natural hazards, policy and fuel-cost shocks, and selected global transmission events, including periods in which momentum-based indicators such as RSI do not provide clear signals. As shown in Section~\ref{sec:results}, volatility flags a substantial share of documented crisis episodes that RSI misses, particularly when price movements reverse within the observation window or when compound shocks produce offsetting directional effects. The workflow is computationally lightweight, does not require model re-estimation as new observations arrive, and is suitable for integration into automated alert systems.

The remainder of the paper proceeds as follows. Section~\ref{sec:data} describes the data source and OHLC structure. Section~\ref{sec:methods} presents the volatility estimators and detection rule. Section~\ref{sec:results} presents the five country case studies. Section~\ref{sec:discussion} discusses implications and limitations. Section~\ref{sec:conclusion} concludes.

\section{Data}\label{sec:data}

\subsection{Price data and OHLC structure}\label{sec:ohlc}

The data are from the World Bank Real-Time Prices (RTP) system, constructed using the methods of \citet{andree2021_inflation} and \citet{andreepape2023}. RTP generates high-frequency food price series by combining price observations from the World Food Programme, FAO, national statistical offices, and other partners with machine-learning-based imputation to fill gaps in space, time, and commodity coverage. The operational system currently covers dozens of fragile and conflict-affected countries, with weekly updates across hundreds of markets and staple commodities. The data have been validated against independent survey and crowdsourcing exercises \citep{adewopo2025} and global commodities data \citep{emediegwu2024}, and have supported country applications in Myanmar \citep{ecker2023}, Afghanistan \citep{gbadegesin2024}, Malawi \citep{kaiyatsa2023}, South Sudan \citep{schincariol2024}, and Sudan \citep{bari2025}. Recent work has also leveraged this data to forecast food-crisis risk \citep{andree2020_predicting_crises,wang2020_stochastic_food_insecurity,wang2022,penson2024}.

At the core of RTP is a fractionally integrated Generalized Autoregressive Conditional Heteroskedasticity (fiGARCH) model with generalized error distribution (GED) innovations applied to partially observed and partially imputed price series, which allows the system to model long-memory volatility dynamics and to track not only the closing price but also synthetic open, high, and low prices within each aggregation period \citep{baillie1996,andree2021_inflation}.\footnote{RTP is implemented as three sub-series: Real-Time Food Prices (RTFP) for food commodities, Real-Time Energy Prices (RTEP) for fuels, and Real-Time Foreign Exchange Rates (RTFX) for currency markets \citep{rtp2021}. The volatility indicators developed here use the RTFP OHLC series, but the methodology extends directly to energy and exchange rate series.} These OHLC estimates are stored at monthly frequency for each market--product combination. While they are model-based rather than directly observed, they summarize within-period price dispersion in a way that is compatible with standard OHLC-based volatility estimators.

The following subsections describe how the synthetic open--high--low--close (OHLC) price series are constructed from a fiGARCH model with GED innovations. The basic idea is to (i)~estimate a long-memory conditional variance model for log returns, (ii)~use the filtered conditional variance and innovation distribution to obtain the expected price range (a central 50\% interval, 25th--75th percentiles) of the conditional price distribution within each period, and (iii)~map this interval into low and high prices while treating the open as the conditional expectation and the close as the realized end-of-period price. Understanding this construction clarifies how the range-based volatility estimators presented in Section~\ref{sec:methods} relate to the underlying data-generating process.

\subsection{fiGARCH--GED model for log returns} \label{sec:fiGARCH-GED}

Let $P_t$ denote the price index at (monthly) time $t$ and $x_t = \ln P_t$
its logarithm. The (log) return over one period is
\begin{equation}
r_t = x_t - x_{t-1} = \ln\left(\frac{P_t}{P_{t-1}}\right).
\end{equation}
In the RTP framework, log returns are modeled as
\begin{equation}
r_t = \mu + \varepsilon_t,
\qquad
\varepsilon_t = \sqrt{h_t}\, z_t,
\end{equation}
where $\mu$ is a constant drift term and $h_t$ is the conditional variance.

Let $\{\mathcal{F}_t\}$ denote the filtration generated by $\{r_s : s \le t\}$. We assume that $\{z_t\}$ is a standardized martingale difference sequence with respect to $\{\mathcal{F}_t\}$, that is
\begin{equation}
\mathbb{E}[z_t \mid \mathcal{F}_{t-1}] = 0,
\qquad
\mathbb{E}[z_t^2 \mid \mathcal{F}_{t-1}] = 1
\quad \text{a.s.}
\end{equation}
Conditional on $\mathcal{F}_{t-1}$ we take $z_t$ to follow a standardized generalized error distribution (GED) with shape parameter $\nu > 0$ and unit variance. Let $F_\nu(\cdot)$ denote the corresponding cumulative distribution function.

The conditional variance process $h_t$ follows a fiGARCH$(1,d,1)$ recursion \citep{baillie1996}. A convenient representation is
\begin{equation}
(1 - \beta_1 L)\, h_t
= \omega
+ \bigl[(1 - \beta_1 L) - (1 - \phi_1 L)(1 - L)^d \bigr] \varepsilon_t^2,
\label{eq:figarch}
\end{equation}
where $L$ is the lag operator $(L h_t = h_{t-1})$, $\omega > 0$, $\beta_1$ and $\phi_1$ are non-negative parameters, and $(1-L)^d$ is the fractional differencing operator with $d \in (0,1)$ capturing long memory in volatility.

Given estimated parameters
\[
\theta = (\mu,\omega,\beta_1,\phi_1,d,\nu),
\]
filtering the model yields sequences of conditional variances $h_t$ and standardized residuals $z_t$ for $t=1,\dots,T$. These in turn define, for each $t$, a conditional distribution for the next-period log return:
\begin{equation}
r_{t+1} \mid \mathcal{F}_t \sim \mu + \sqrt{h_{t+1}}\, z_{t+1},
\qquad
z_{t+1} \sim \text{GED}(0,1,\nu),
\end{equation}
where $\mathcal{F}_t$ denotes the information set up to time $t$.

\subsubsection{Generalized error distribution and tail behavior}
\label{sec:ged}

The GED allows the conditional variance $h_t$ and the tail thickness (kurtosis) of returns to be modeled separately. The probability density function (pdf) of a GED with location $0$, scale parameter $\beta > 0$ and shape $\nu$ is
\begin{equation}
f_{\text{GED}}(z; \beta, \nu)
= \frac{\nu}{2 \beta \,\Gamma(1/\nu)}
  \exp\!\left\{-\left(\frac{|z|}{\beta}\right)^{\nu}\right\},
\qquad z \in \mathbb{R},
\end{equation}
where $\Gamma(\cdot)$ is the gamma function. The variance of this distribution is
\begin{equation}
\operatorname{Var}(z) = \beta^2 \,\frac{\Gamma(3/\nu)}{\Gamma(1/\nu)}.
\end{equation}
To obtain standardized innovations with unit variance, we set
\begin{equation}
\beta(\nu)
= \left(\frac{\Gamma(1/\nu)}{\Gamma(3/\nu)}\right)^{1/2},
\end{equation}
and write
\begin{equation}
z_t \sim \text{GED}\bigl(0,\,\beta(\nu),\,\nu\bigr),
\qquad
\mathbb{E}[z_t] = 0,
\quad
\operatorname{Var}(z_t) = 1.
\end{equation}

The kurtosis of the standardized GED is
\begin{equation}
\kappa(\nu)
= \frac{\mathbb{E}[z_t^4]}{\{\operatorname{Var}(z_t)\}^2}
= \frac{\Gamma(5/\nu)\,\Gamma(1/\nu)}{\Gamma(3/\nu)^2}.
\end{equation}
When $\nu = 2$, the GED reduces to the Gaussian distribution and $\kappa(2) = 3$. For $\nu < 2$, the distribution is leptokurtic with heavier tails and $\kappa(\nu) > 3$, implying a higher probability of large realizations of $|z_t|$ (and thus of large returns) relative to the Gaussian case. For $\nu > 2$, the tails become lighter than Gaussian.

In the fiGARCH--GED model, the conditional variance $h_t$ and the shape parameter $\nu$ are estimated jointly. This separation has two important implications for the behavior of the OHLC series and the range-based volatility estimators:

\begin{enumerate}
    \item \textbf{Sharp but temporary price swings.} Because the innovations are heavy-tailed when $\nu < 2$, a single large ``candle'' (i.e., an unusually large movement in the price index) can be accommodated by a large realization of $z_t$ without implying that all subsequent innovations must also be large. In the fiGARCH recursion~\eqref{eq:figarch}, a large squared shock $\varepsilon_t^2 = h_t z_t^2$ will enter the variance dynamics, but its impact on future $h_{t+k}$ depends on both the long-memory structure (through $d$ and the lag polynomials) and the frequency of such shocks. A single extreme return can thus be interpreted partly as a tail event in $z_t$ rather than as a permanent shift in the volatility regime.
    \item \textbf{Persistent volatility regimes.} At the same time, the fractional differencing parameter $d \in (0,1)$ and the autoregressive structure in~\eqref{eq:figarch} allow $h_t$ to exhibit long memory. When a sequence of large shocks occurs (for example, during a war, siege, or macroeconomic crisis), the fiGARCH recursion will propagate elevated $\varepsilon_t^2$ forward, generating a persistent high-volatility regime. In this case, the GED tails and the long-memory variance dynamics work together: the GED captures the excess kurtosis within the regime, while fiGARCH ensures that the regime itself is persistent.
\end{enumerate}

\subsubsection{Conditional distribution of log prices and central 50\% spread}

The one-step-ahead conditional log price is
\begin{equation}
x_{t+1} = x_t + r_{t+1}.
\end{equation}
Given $\mathcal{F}_t$, its conditional distribution is
\begin{equation}
x_{t+1} \mid \mathcal{F}_t \sim m_{t+1} + \sqrt{h_{t+1}}\, z_{t+1},
\qquad
m_{t+1} = x_t + \mu,
\end{equation}
with $z_{t+1}$ distributed as in Section~\ref{sec:ged}.

Let $F_\nu^{-1}(\cdot)$ denote the quantile function of the standardized GED. For any probability level $p \in (0,1)$, the conditional $p$-quantile of $x_{t+1}$ is
\begin{equation}
q_{p,t+1}
= m_{t+1} + \sqrt{h_{t+1}}\, a_p(\nu),
\qquad
a_p(\nu) = F_\nu^{-1}(p).
\end{equation}

We are particularly interested in the central 50\% interval, defined by the 25th and 75th percentiles of the conditional log-price distribution:
\begin{equation}
q_{0.25,t+1} = m_{t+1} + \sqrt{h_{t+1}}\, a_{0.25}(\nu),
\qquad
q_{0.75,t+1} = m_{t+1} + \sqrt{h_{t+1}}\, a_{0.75}(\nu).
\end{equation}
Equivalently, these satisfy the probability-mass conditions
\begin{equation}
\mathbb{P}\bigl(x_{t+1} \le q_{0.25,t+1} \mid \mathcal{F}_t\bigr) = 0.25,
\qquad
\mathbb{P}\bigl(x_{t+1} \le q_{0.75,t+1} \mid \mathcal{F}_t\bigr) = 0.75.
\end{equation}
They can be viewed as central-interval analogues of the tail integrals used in expected shortfall \citep{acerbitasche2002} calculations: the math is identical but instead of integrating probability mass in the tails, we delimit the central 50\% of the conditional distribution.

Mapping these quantiles back to the price level gives the model-based low and high candidates,
\begin{equation}
L^{*}_{t+1} = \exp\bigl(q_{0.25,t+1}\bigr),
\qquad
H^{*}_{t+1} = \exp\bigl(q_{0.75,t+1}\bigr).
\end{equation}

The distance between these quantiles,
\begin{equation}
q_{0.75,t+1} - q_{0.25,t+1}
= \sqrt{h_{t+1}}\bigl[a_{0.75}(\nu) - a_{0.25}(\nu)\bigr],
\end{equation}
is increasing in both $\sqrt{h_{t+1}}$ and the tail thickness implied by $\nu$ (since the GED quantiles move outward as the distribution becomes more heavy-tailed).

\subsubsection{Definition of open, high, low and close}

For each period $t+1$, we construct OHLC prices as follows.

\paragraph{Close.} The close is taken to be the realized end-of-period price index from the partially imputed price survey system:
\begin{equation}
C_{t+1} = P_{t+1} = \exp(x_{t+1}^{\text{obs}}),
\end{equation}
where $x_{t+1}^{\text{obs}}$ is the end-of-month log price.

\paragraph{Open.} The open is set equal to the conditional expectation of the log price, mapped back to levels:
\begin{equation}
O_{t+1}
= \exp\bigl( \mathbb{E}[x_{t+1} \mid \mathcal{F}_t] \bigr)
= \exp(m_{t+1}).
\end{equation}

\paragraph{Model-based low and high.} Using the conditional quantiles defined above, we obtain the central 50\% price interval:
\begin{equation}
L^{*}_{t+1} = \exp\bigl(m_{t+1} + \sqrt{h_{t+1}}\, a_{0.25}(\nu)\bigr),
\qquad
H^{*}_{t+1} = \exp\bigl(m_{t+1} + \sqrt{h_{t+1}}\, a_{0.75}(\nu)\bigr).
\end{equation}

\paragraph{Final high and low.} To ensure that the realized open and close lie within the range, and that the range always spans the most relevant parts of the modeled distribution, the final high and low are defined as
\begin{equation}
H_{t+1} = \max\{\,O_{t+1},\,C_{t+1},\,H^{*}_{t+1}\,\},
\qquad
L_{t+1} = \min\{\,O_{t+1},\,C_{t+1},\,L^{*}_{t+1}\,\}.
\end{equation}

This construction guarantees that
\begin{equation}
L_{t+1} \le \min\{O_{t+1},C_{t+1}\}
\quad\text{and}\quad
H_{t+1} \ge \max\{O_{t+1},C_{t+1}\},
\end{equation}
and that, absent extreme realizations of $C_{t+1}$, the interval $[L_{t+1},H_{t+1}]$ is anchored around the central 50\% of the conditional price distribution implied by the fiGARCH--GED model.

\section{Methods}\label{sec:methods}

Range-based volatility estimators exploit the information contained in intra-period price extremes. Under standard diffusion models, these estimators achieve substantial efficiency gains relative to close-to-close measures, sometimes by an order of magnitude \citep{parkinson1980,garmanklass1980,yangzhang2000}. This section presents the volatility estimators, describes their relationship to the OHLC data structure, and specifies the procedure used to identify elevated volatility episodes in the empirical application.

\subsection{Setup and notation}

We observe prices at equally spaced times $(t = 0, 1, 2, \dots)$. For each observation period $(t)$ we denote:
\begin{itemize}
    \item $O_t$: opening price
    \item $H_t$: highest price during the period
    \item $L_t$: lowest price during the period
    \item $C_t$: closing price
\end{itemize}

We work with natural logarithms of price ratios. For brevity, define:
\begin{itemize}
    \item Log range: $d_t = \ln(H_t / L_t)$
    \item Log open-close return: $c_t = \ln(C_t / O_t)$
    \item ``Overnight'' log return (close-to-open between periods): $o_t = \ln(O_t / C_{t-1})$
\end{itemize}

As a continuous-time benchmark we assume that the efficient price follows a geometric Brownian motion,
\begin{equation}
dP_t = \mu P_t \, dt + \sigma P_t \, dW_t,
\end{equation}
so that log prices satisfy
\begin{equation}
d\ln P_t = \left(\mu - \tfrac{1}{2}\sigma^2\right)dt + \sigma dW_t.
\end{equation}

Let $\Delta t$ be the length of one observation period in years (e.g., $\Delta t = 1/252$ for daily data, $\Delta t = 1/12$ for monthly data). Then $\sigma^2 \Delta t$ is the per-period variance parameter. The estimators below aim to estimate this per-period variance and then annualize it by multiplying by an annualization factor $N \approx 1 / \Delta t$ and taking a square root.

All estimators are computed over rolling windows of length $n$ periods. For a generic per-period variance estimator $\hat{v}_t$ over a window ending at time $t$, the corresponding annualized volatility is
\begin{equation}
\hat{\sigma}_t = \sqrt{N \,\hat{v}_t}.
\end{equation}

\subsection{Volatility Estimators}

The following definitions present the volatility estimators used in this paper. All estimators follow the same pattern: a per-period variance component is estimated from the OHLC data over a rolling window, annualized by multiplying by an annualization factor $N$, and converted to volatility by taking a square root.

\begin{definition}[Close-to-close volatility]
The close-to-close estimator uses only closing prices and treats the returns as arithmetic rates of change,
\[
r_t = \frac{C_t - C_{t-1}}{C_{t-1}}.
\]
Given a window of $m$ returns $\{r_i\}_{i = t-m+1}^{t}$, the sample variance is
\[
\widehat{\operatorname{Var}}(r)_t
= \frac{1}{m-1} \sum_{i = t-m+1}^{t} \bigl(r_i - \bar{r}_t\bigr)^2,
\qquad
\bar{r}_t = \frac{1}{m} \sum_{i = t-m+1}^{t} r_i.
\]
The annualized close-to-close volatility is then
\[
\hat{\sigma}_{\mathrm{CC},t} = \sqrt{N \,\widehat{\operatorname{Var}}(r)_t}.
\]
In the implementation we use $m = n-1$ returns corresponding to $n$ prices. For small period-to-period changes, arithmetic returns and log returns are numerically very close; all range-based estimators below use log price ratios exactly.
\end{definition}

\begin{definition}[Parkinson high-low estimator]
Parkinson's estimator uses only the high and low within each period. For each $t$, define the log range $d_t = \ln(H_t / L_t)$. Under the geometric Brownian motion model with zero drift over the period and continuous trading, the log range satisfies the moment condition
\[
\mathbb{E}\bigl[d_t^2\bigr]
= 4 \ln 2 \,\sigma^2 \Delta t.
\]
A method-of-moments estimator of the per-period variance based on this identity is
\[
\hat{v}_{P,t}
= \frac{1}{4 n \ln 2}
\sum_{i = t-n+1}^{t} d_i^2,
\]
and the corresponding annualized Parkinson volatility is $\hat{\sigma}_{P,t} = \sqrt{N \,\hat{v}_{P,t}}$.
This estimator is substantially more efficient than close-to-close when the zero-drift, continuous-trading assumptions are reasonable, but it can become biased in strongly trending markets.
\end{definition}

\begin{definition}[Garman-Klass estimator]
\citet{garmanklass1980} derive an estimator that uses the range and the open-close return. Using $d_t = \ln(H_t / L_t)$ and $c_t = \ln(C_t / O_t)$, their per-period variance estimator can be written as
\[
\hat{v}_{GK,t} = \frac{1}{n} \sum_{i = t-n+1}^{t} \left(\frac{1}{2} d_i^2 - (2\ln 2 - 1)\, c_i^2 \right).
\]
Under geometric Brownian motion with zero drift within each period, this is an unbiased estimator of the per-period variance $\sigma^2 \Delta t$, and is more efficient than both close-to-close and Parkinson's range estimator. The corresponding annualized volatility is $\hat{\sigma}_{GK,t} = \sqrt{N \,\hat{v}_{GK,t}}$.
\end{definition}

\begin{definition}[Rogers-Satchell estimator]
\citet{rogerssatchell1991} propose an estimator that remains unbiased for variance even in the presence of non-zero drift. Let $u_t = \ln(H_t / O_t)$, $\ell_t = \ln(L_t / O_t)$, and $c_t = \ln(C_t / O_t)$. One convenient equivalent form of the Rogers-Satchell contribution for period $t$ is
\[
q_t
= \ln\left(\frac{H_t}{C_t}\right)\ln\left(\frac{H_t}{O_t}\right)
+
\ln\left(\frac{L_t}{C_t}\right)\ln\left(\frac{L_t}{O_t}\right).
\]
The per-period variance estimator over a window is $\hat{v}_{RS,t} = \frac{1}{n} \sum_{i = t-n+1}^{t} q_i$, and the annualized Rogers-Satchell volatility is $\hat{\sigma}_{RS,t} = \sqrt{N \,\hat{v}_{RS,t}}$.
Under the geometric Brownian motion model with arbitrary constant drift $\mu$, this estimator is unbiased for $\sigma^2 \Delta t$, making it more robust than Garman-Klass when there are strong trends.
\end{definition}

\begin{definition}[Garman-Klass-Yang-Zhang extension]
The Garman-Klass-Yang-Zhang (``GK-YZ'') extension \citep{yangzhang2000} augments the Garman-Klass estimator by incorporating squared close-to-open jumps between periods. Define the ``overnight'' log return $o_t = \ln(O_t / C_{t-1})$, and retain $d_t$ and $c_t$ as before. The GK-YZ per-period variance estimator over a window is
\[
\hat{v}_{GKYZ,t} = \frac{1}{n} \sum_{i = t-n+1}^{t} \left[ o_i^2 + \frac{1}{2} d_i^2 - (2\ln 2 - 1)\, c_i^2 \right],
\]
with annualized volatility $\hat{\sigma}_{GKYZ,t} = \sqrt{N \,\hat{v}_{GKYZ,t}}$.
When between-period jumps (e.g., market closes and reopens) are important, this estimator captures both those jumps and within-period variation.
\end{definition}

\begin{definition}[Yang-Zhang estimator]
\citet{yangzhang2000} propose a volatility estimator that combines three components: (i)~the variance of close-to-open returns $o_t$, (ii)~the variance of open-close returns $c_t$, and (iii)~the intraday Rogers-Satchell variance.
The close-to-open and open-close components are computed as sample variances over the last $n$ periods:
\[
\hat{v}_{o,t}
= \widehat{\operatorname{Var}}(o_i)_{i = t-n+1}^{t},
\qquad
\hat{v}_{c,t}
= \widehat{\operatorname{Var}}(c_i)_{i = t-n+1}^{t}.
\]
The Rogers-Satchell component $\hat{v}_{RS,t}$ is as defined above. The Yang-Zhang per-period variance estimator is a weighted sum,
\[
\hat{v}_{YZ,t} = \hat{v}_{o,t} + k \,\hat{v}_{c,t} + (1 - k)\,\hat{v}_{RS,t},
\]
with weight $k \in (0,1)$ chosen to balance the contributions of the open-close and Rogers-Satchell components. In the implementation we use
\[
k
= \frac{\alpha - 1}{\alpha + \frac{n+1}{n-1}},
\]
with $\alpha \approx 1.34$. The annualized Yang-Zhang volatility is $\hat{\sigma}_{YZ,t} = \sqrt{N \,\hat{v}_{YZ,t}}$.
The Yang-Zhang estimator is designed to be drift-independent (through the use of the Rogers-Satchell intraday component) and to incorporate between-period jumps efficiently. In empirical applications it typically exhibits lower variance than simpler historical and range-based estimators \citep{yangzhang2000}.
\end{definition}

\subsection{Relationship to the OHLC data structure}\label{sec:OHLC_relation}

Given the constructed OHLC series $\{O_t,H_t,L_t,C_t\}$ described in Section~\ref{sec:ohlc}, the range-based estimators can be interpreted as functions of the modeled conditional distribution over rolling windows. For each period $t$, the key logarithmic components---$d_t = \ln(H_t/L_t)$, $c_t = \ln(C_t/O_t)$, and $o_t = \ln(O_t/C_{t-1})$---summarize, in different ways, the conditional spread and location of the modeled price distribution within each period.

Under the fiGARCH--GED specification, these quantities inherit structure from the underlying volatility process. Over a window of length $n$, the range-based estimators use averages of functions of $\{d_i, c_i, o_i\}_{i=t-n+1}^{t}$ to estimate per-period variance, which is then annualized. For example, the Parkinson estimator uses the average of $d_i^2$, while Garman--Klass and Rogers--Satchell combine $d_i^2$ with $c_i^2$ and other log differences. In our setting, these estimators summarize how the modeled conditional distribution (through $h_t$ and the GED quantiles) evolves over time, with the OHLC mapping providing a coherent bridge between the latent volatility process and observable price indices.

A key property of this setup is that the range-based estimators remain well-defined and informative even though the OHLC series are model-based rather than directly observed. The theoretical efficiency gains of range-based estimators derive from their use of within-period price extremes to reduce variance in volatility estimation. In the RTP framework, these extremes are constructed from the conditional distribution implied by the fiGARCH--GED model, which captures both the level of conditional variance ($h_t$) and the tail behavior of innovations (through the GED shape parameter $\nu$). The resulting high and low prices thus encode information about within-period dispersion that is consistent with the underlying volatility dynamics.

It is instructive to consider limiting cases in which the additional information in the range becomes negligible and range-based estimators effectively collapse to standard close-to-close volatility measures. These cases clarify the conditions under which the range-based approach provides meaningful incremental information.

\begin{enumerate}
    \item \textbf{Vanishing conditional variance.} If $h_t \to 0$ for all $t$ (for example, in a hypothetical perfectly stable market), then $q_{0.25,t+1} \approx q_{0.75,t+1} \approx m_{t+1}$, so that $L^{*}_{t+1} \approx H^{*}_{t+1} \approx \exp(m_{t+1})$. In the absence of large deviations of $C_{t+1}$ from its expectation, we obtain $H_{t+1} \approx L_{t+1} \approx O_{t+1} \approx C_{t+1}$ and thus $d_t = \ln(H_t/L_t) \approx 0$ for all $t$. In this limit, range-based estimators contribute little beyond the information in the close-to-close returns $r_t$.
    \item \textbf{Open and close dominate the range.} Even when $h_t$ is not negligible, if the realized close $C_t$ and the conditional open $O_t$ lie well inside the modeled central interval, then $H_t$ and $L_t$ will typically be close to $H^{*}_t$ and $L^{*}_t$. Conversely, if $C_t$ and $O_t$ coincide or nearly coincide (for example, in periods with very small net movement), the max/min definitions may again yield $H_t \approx L_t$, shrinking $d_t$. In this case, range-based estimators become numerically similar to close-to-close estimators computed from $r_t$ and $c_t$.
    \item \textbf{Degenerate OHLC construction.} As an extreme thought experiment, if we were to set $H_t = L_t = C_t$ for all $t$, then $d_t = \ln(H_t/L_t) = 0$ identically, and all range-based estimators reduce to functions of close-to-close (or open--close) returns alone. They no longer encode any within-period dispersion.
\end{enumerate}

In practice, the fractionally integrated GARCH model with GED innovations routinely generates nontrivial conditional variance $h_t$ in crisis-prone environments, and the OHLC mapping preserves a meaningful spread between $L_t$ and $H_t$. In these settings, range-based volatility estimators exploit the additional information contained in the model-implied highs and lows. This yields more efficient volatility measurement than close-to-close alternatives, while remaining straightforward to compute in an operational workflow. The empirical results in Section~\ref{sec:results} confirm that the range-based estimators produce meaningfully different signals than close-to-close measures, consistent with the OHLC structure encoding substantive within-period dispersion information.

\subsection{Detection rule and operational implementation}\label{sec:detection}

In our empirical application, all volatility estimators are computed with a rolling window of length $n=10$, and annualization assumes $\Delta t = 1/12$, so that $N = 12$ in all results. For operational use we employ a simple volatility shock indicator based on two threshold conditions:
\begin{enumerate}
    \item \textbf{Relative threshold:} volatility above its 12-month simple moving average,
    \item \textbf{Tail threshold:} volatility above a high percentile (the 85th percentile) of its historical distribution computed over the full sample, and not below a low percentile (the 20th percentile).
\end{enumerate}
Months satisfying both conditions are flagged as volatility episodes.

The percentile thresholds are not highly sensitive parameters: the relative-threshold condition (above the 12-month moving average) does most of the work in identifying departures from recent baseline conditions, while the percentile bounds serve primarily as guardrails. The upper bound ensures that episodes remain flagged during elevated dispersion even as volatility recedes from its peak, avoiding false negatives during highly disrupted periods. The lower bound prevents the indicator from triggering during sustained low-volatility regimes, where minor upticks may exceed the moving average without representing meaningful stress. Using sample-wide percentiles also allows the early part of the series (including the 2007--08 global food price crisis, which falls within the initial rolling window) to be evaluated against a consistent historical benchmark. Varying the percentile thresholds within reasonable ranges (e.g., 80th--90th for the upper bound, 15th--25th for the lower bound) does not materially alter the set of flagged episodes in the applications presented below.

This rule is simple, transparent, and easy to implement in operational dashboards alongside existing indicators such as RSI and MACD \citep{appel1979,wilder1978}. The percentile thresholds provide straightforward calibration parameters that can be tuned against historical event data. The rule is intentionally minimal: even with default settings, volatility flags episodes that momentum indicators miss.

\section{Results}\label{sec:results}

This section presents five case studies that assess OHLC-based volatility estimators against documented shocks in remote, data-scarce settings spanning Sub-Saharan Africa, the Middle East, the Caribbean, and Southeast Asia. The applications cover diverse shock types (conflict, natural hazards, policy change, global transmission) and market structures (landlocked versus coastal, import-dependent versus producer). In each case, the RTP food price index is used and flagged volatility episodes are interpreted against event timelines compiled from the literature (Appendix). Applications are presented alphabetically.

\subsection{Al Fashir, Sudan (2007--2025)}\label{sec:sudan_case}

\begin{figure}[H]
    \centering
    \includegraphics[width=0.925\textwidth]{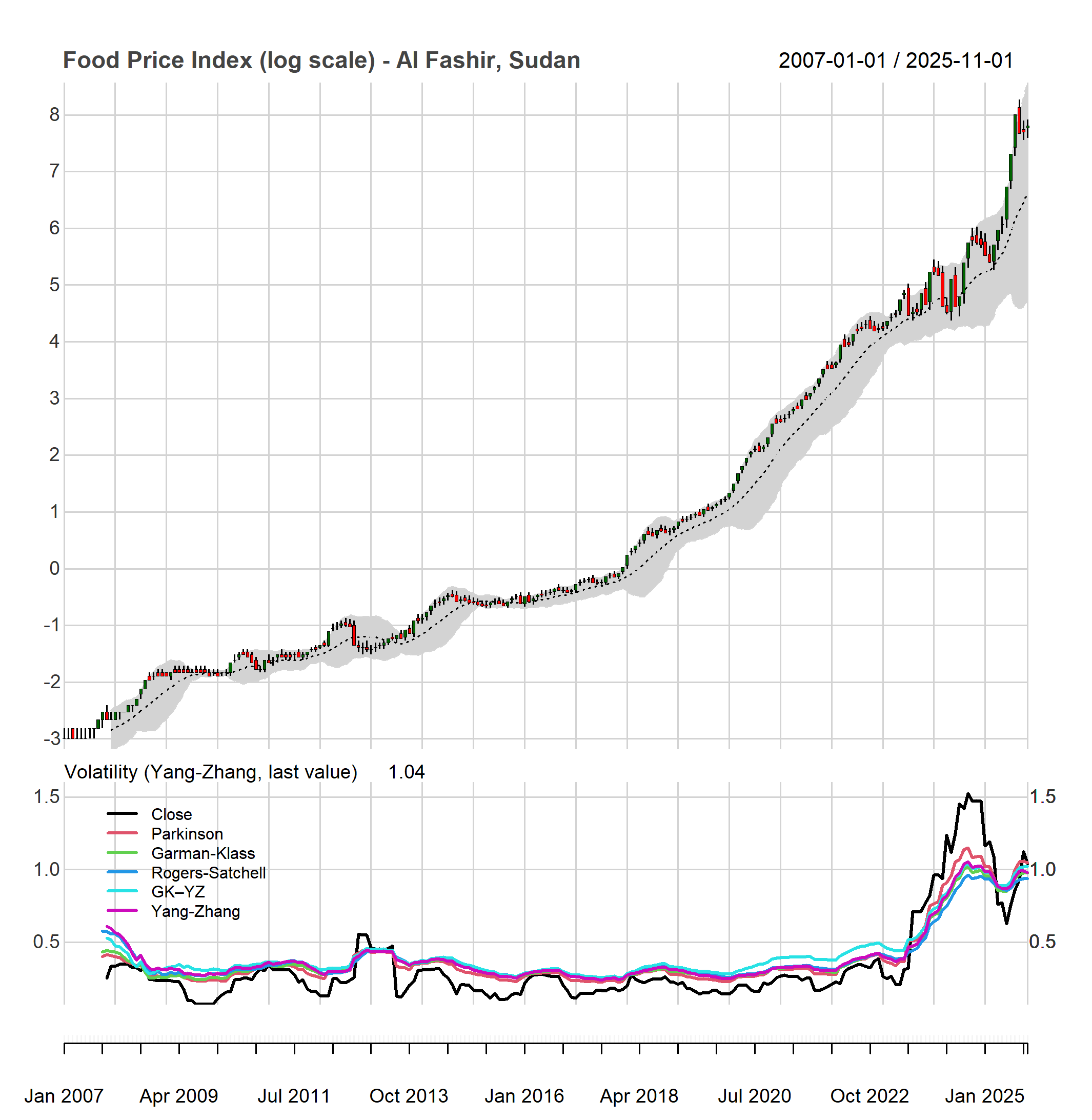}
    \caption{Open, High, Low, Close food prices (index, log) in Al Fashir, Sudan, plotted on a candlestick chart together with six OHLC-based volatility metrics. In the candlestick chart (top panel), each bar represents one month: the vertical line spans the low to high, while the box spans open to close; filled (red) boxes indicate months where the close fell below the open, and hollow (green) boxes indicate months where the close exceeded the open. The dotted line shows the 12-month moving average, and the grey shading indicates Bollinger bands (two standard deviations around the moving average). Subsequent panels show close-to-close volatility and five range-based estimators (Parkinson, Garman--Klass, Rogers--Satchell, Garman--Klass--Yang--Zhang, and Yang--Zhang). All estimators identify the same broad stress periods but differ in smoothness: range-based measures respond more quickly to widening intra-period dispersion, while close-to-close volatility reacts only to month-end price jumps.}
    \label{fig:vol_all}
\end{figure}

Sudan is a demanding test case because global shocks, macroeconomic instability, and recurrent conflict overlap over long horizons, with severe impacts in Darfur. We focus on Al Fashir, the capital of North Darfur and a regional trading hub surrounded by major IDP camps including Zamzam, Abu Shouk, and Al Salam. Figure~\ref{fig:vol_all} shows the six volatility estimators computed from the local OHLC series; Appendix~\ref{app:additional_figures} presents the same figure for the other cases.

All estimators detect the main stress periods but differ in smoothness and sensitivity to within-period dispersion. Range-based measures rise earlier and remain elevated when intra-month ranges widen persistently, whereas close-to-close volatility reacts sharply to month-end jumps but can understate stress when prices fluctuate without a clear directional move. The Yang--Zhang estimator offers a practical compromise, tracking the more responsive range-based metrics while reducing noise from drift and opening-price discontinuities: it incorporates the drift-independent Rogers--Satchell component and weights the open-close and overnight components to balance responsiveness against variance. We therefore use Yang--Zhang volatility, $\hat{\sigma}_{YZ,t}$, as the primary measure below. Figure~\ref{fig:vol_signal} shows the main result; the event timeline is in Appendix~\ref{app:sudan_events}.

The bottom panel plots $\hat{\sigma}_{YZ,t}$ and its 12-month moving average; red segments mark elevated volatility. The signals align with major disruptions, including the 2007--08 global food price crisis, the post-2011 adjustment period, the 2018--2019 bread-protests and political transition, the 2021 coup, and the SAF--RSF war from April 2023. RSI remains elevated for much of the period, consistent with chronic inflation, but does not isolate acute stress and even declines during severe disruptions. Volatility, by contrast, consistently flags conflict escalation, policy shocks, and market fragmentation. The divergence is most pronounced after April 2023, when RSI moderates while volatility signals sharp deterioration. Al Fashir illustrates how dispersion-based monitoring complements price-level indicators by separating persistent inflation from episodic market dysfunction.

\begin{figure}[H]
    \centering
    \includegraphics[width=0.75\textwidth]{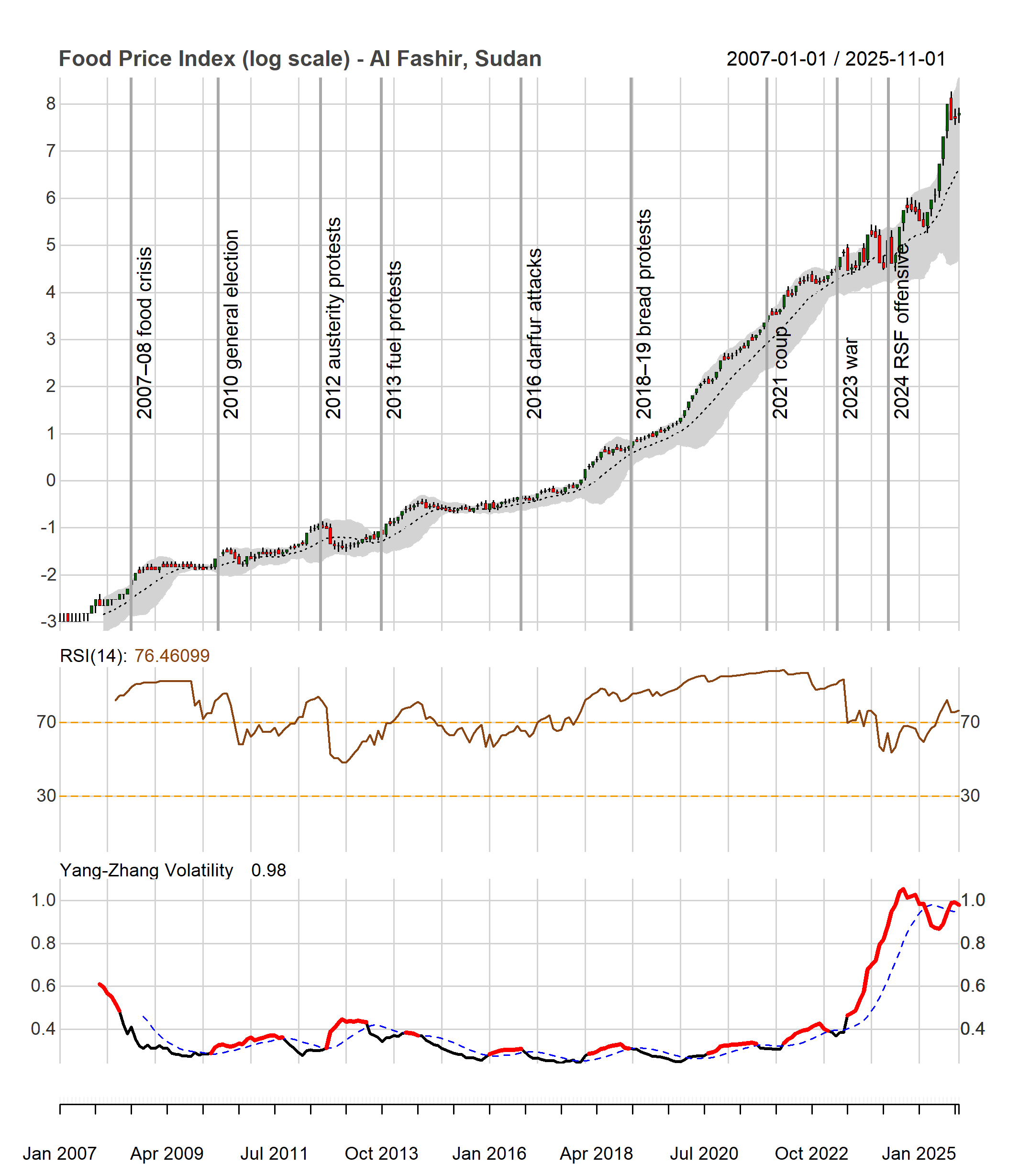}
    \caption{Technical setup to detect volatility shocks. Top: log food prices in Al Fashir, Sudan. Middle: RSI with 30 and 70 reference levels. Bottom: Yang--Zhang annualized volatility, its 12-month moving average, and detected high-volatility episodes (red segments).}
    \label{fig:vol_signal}
\end{figure}

\subsection{Baidoa, Somalia (2007--2025)}\label{sec:somalia_case}

Somalia provides a stringent test case because market conditions are repeatedly stressed by compound climate shocks, protracted insecurity, and episodic disruptions to trade and humanitarian access. We focus on Baidoa, the capital of South West State (Bay Region), a major agricultural market and relief hub highly exposed to rainfall variability and transport constraints. Figure~\ref{fig:som_signal} shows the result; the event timeline is provided in Appendix~\ref{app:somalia_events}.

Volatility rises sharply during well-documented crisis windows, including the 2011 famine, renewed stress in 2014--2015 linked to poor seasonal performance and the 2016--2017 drought emergency. Later clusters coincide with COVID-19 disruptions and the concurrent desert locust upsurge in 2020, as well as the 2023--2024 Deyr and Gu flood episodes. RSI provides some signal during the 2011 crisis and the 2016--2017 drought, but remains within the 30--70 neutral band during several documented stress periods, including the 2020 compound disruption and the 2023--2024 floods, when volatility clearly flags elevated dispersion. Overall, Baidoa illustrates that volatility responds strongly during compound climate and insecurity episodes that fragment markets and disrupt supply.

\begin{figure}[htbp]
    \centering
    \includegraphics[width=0.75\textwidth]{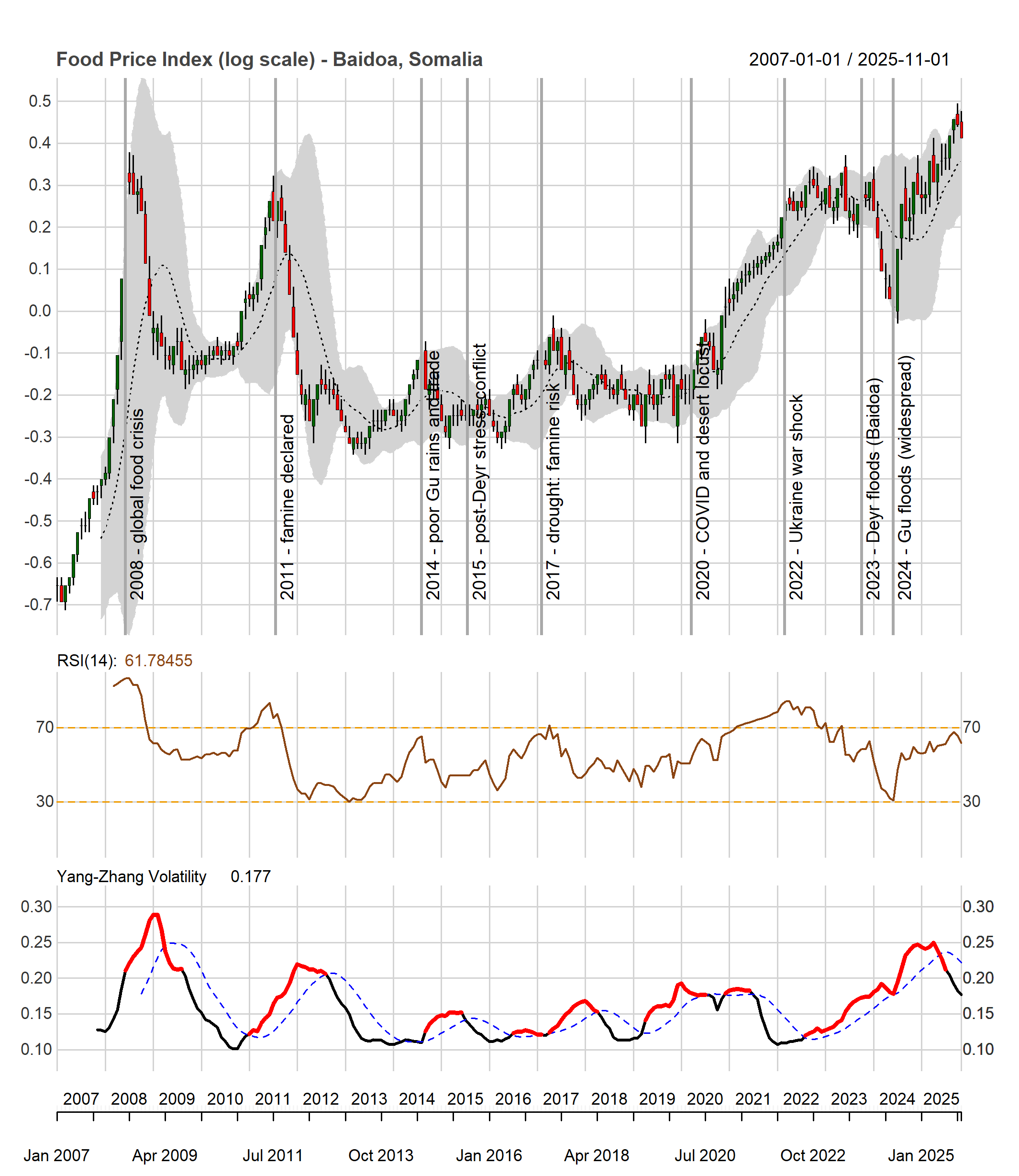}
    \caption{Somalia (Baidoa): technical setup and detected stress episodes. Top panel: log food prices. Middle panel: RSI with 30 and 70 reference levels. Bottom panel: Yang--Zhang annualized volatility, its 12-month moving average, and detected high-volatility episodes (red segments).}
    \label{fig:som_signal}
\end{figure}

\subsection{Far North Region, Cameroon (2010--2025)}\label{sec:cameroon_case}

Cameroon's Far North provides a useful test case because food markets are repeatedly exposed to overlapping security, climate, and policy shocks in a border-sensitive economy. The region lies at the intersection of trade corridors with Nigeria and Chad and has faced sustained disruption from insurgency spillovers, transport constraints, and recurrent floods that periodically impair market access. The setting is also sensitive to cross-border frictions and cost transmission, with a clear shift in volatility after 2022. Figure~\ref{fig:cmr_signal} shows the result; the event timeline is provided in Appendix~\ref{app:cameroon_events}.

The volatility signal shows multiple stress episodes consistent with insecurity, flood-related disruption, cross-border trade constraints, and cost shocks transmitted through transport and distribution margins. RSI remains within the 30--70 band for much of the period and does not register several salient disruptions, whereas volatility clearly flags episodes such as the 2017 floods and renewed attacks in 2018. After 2022 the series shifts into a higher-volatility regime driven by repeated sharp price spikes and reversals. These movements leave RSI relatively unresponsive---reflecting offsetting directional movements within the observation window---but are captured by sustained increases in dispersion and repeated flagged months. Overall, the Far North case illustrates that volatility captures both local fragmentation risks and externally driven cost transmission in a setting where logistics and border dynamics shape price formation.

\begin{figure}[htbp]
    \centering
    \includegraphics[width=0.75\textwidth]{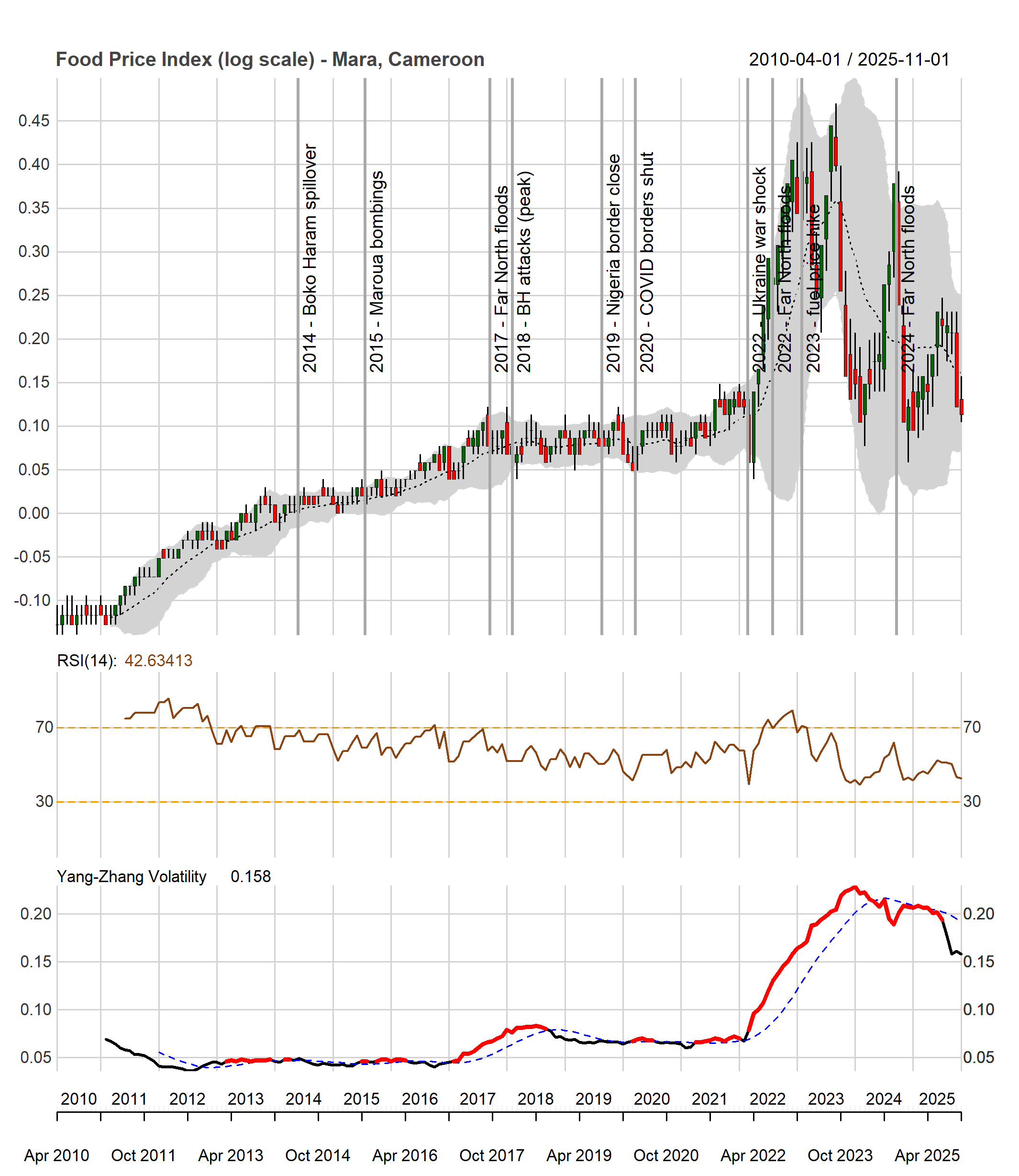}
    \caption{Cameroon (Far North): technical setup and detected stress episodes. Top panel: log food prices. Middle panel: RSI with 30 and 70 reference levels. Bottom panel: Yang--Zhang annualized volatility, its 12-month moving average, and detected high-volatility episodes (red segments).}
    \label{fig:cmr_signal}
\end{figure}

\subsection{Port-au-Prince, Haiti (2007--2025)}\label{sec:haiti_case}

Haiti provides a distinct test case in which volatility is driven by large, discrete disruptions in an urban, import-dependent market that concentrates political risk, infrastructure fragility, and insecurity. Port-au-Prince anchors national logistics and trade and has experienced repeated shocks that generate abrupt breaks in market functioning, including natural disasters, political disruption, fuel-access crises, and worsening security conditions. Figure~\ref{fig:hti_signal} shows the results; the event timeline is provided in Appendix~\ref{app:haiti_events}.

\begin{figure}[htbp]
    \centering
    \includegraphics[width=0.75\textwidth]{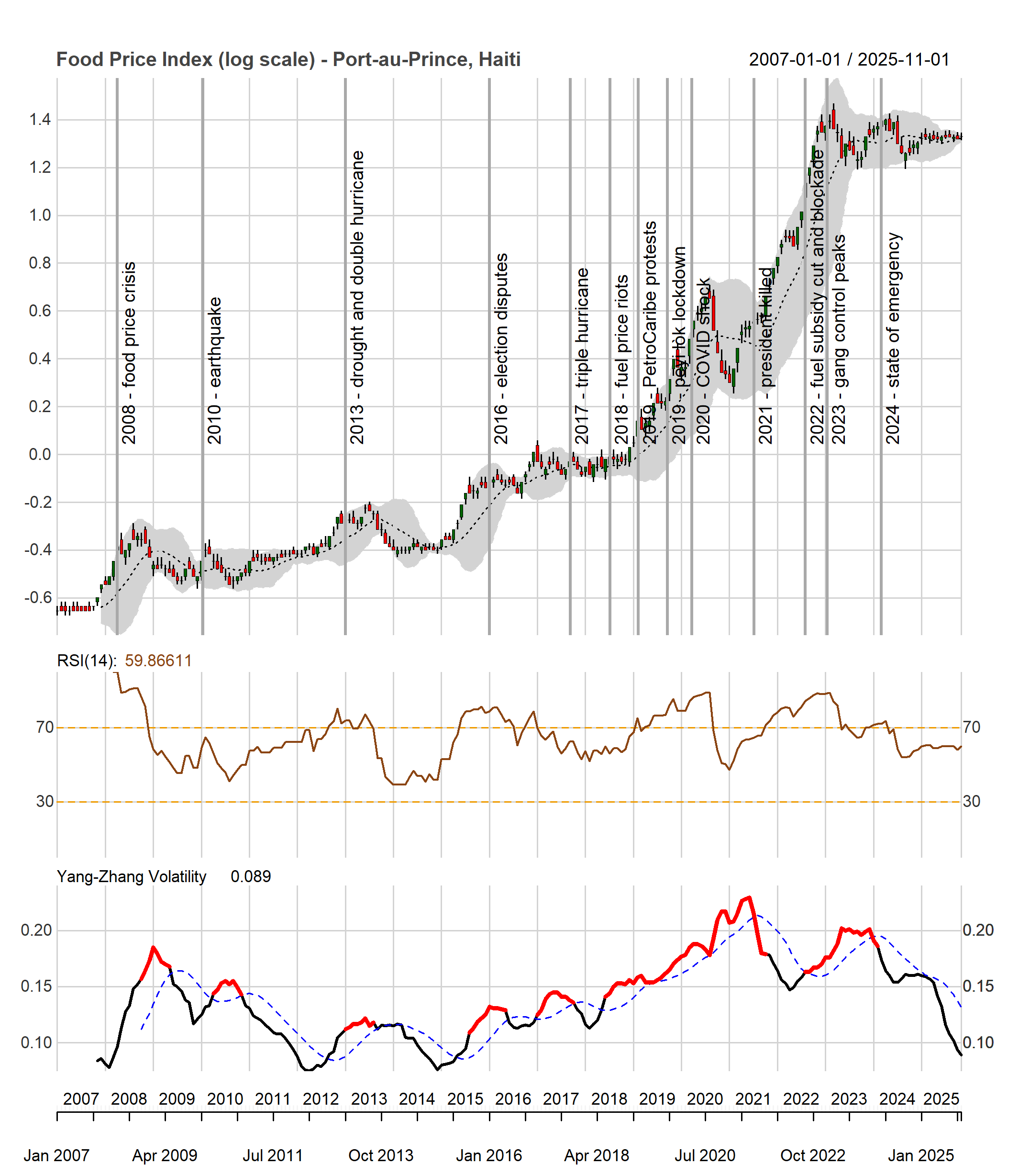}
    \caption{Haiti (Port-au-Prince): technical setup and detected stress episodes. Top panel: log food prices. Middle panel: RSI with 30 and 70 reference levels. Bottom panel: Yang--Zhang annualized volatility, its 12-month moving average, and detected high-volatility episodes (red segments).}
    \label{fig:hti_signal}
\end{figure}

 Volatility rises sharply around major disaster and political-disruption windows and remains elevated during later fuel and logistics shocks. The series shows strong volatility during the 2007--08 crisis and the 2010 earthquake, with renewed clusters during later disruption windows including the 2018 fuel riots, the extended \emph{peyi l\`ok} period in 2019, and the 2022 fuel-access crisis associated with gang control of key infrastructure and transport corridors. Volatility remains elevated through the subsequent deterioration in security conditions. RSI and volatility sometimes co-move during price run-ups, but volatility more consistently captures disaster impacts and market dysfunction---notably the 2010 earthquake and the 2017 hurricane season---when RSI does not breach standard threshold levels.

\subsection{Sulu, Philippines (2007--2025)}\label{sec:philippines_case}

Figure~\ref{fig:phl_signal} shows the detection setup in the Philippines; the event timeline is provided in Appendix~\ref{app:philippines_events}.

\begin{figure}[htbp]
    \centering
    \includegraphics[width=0.75\textwidth]{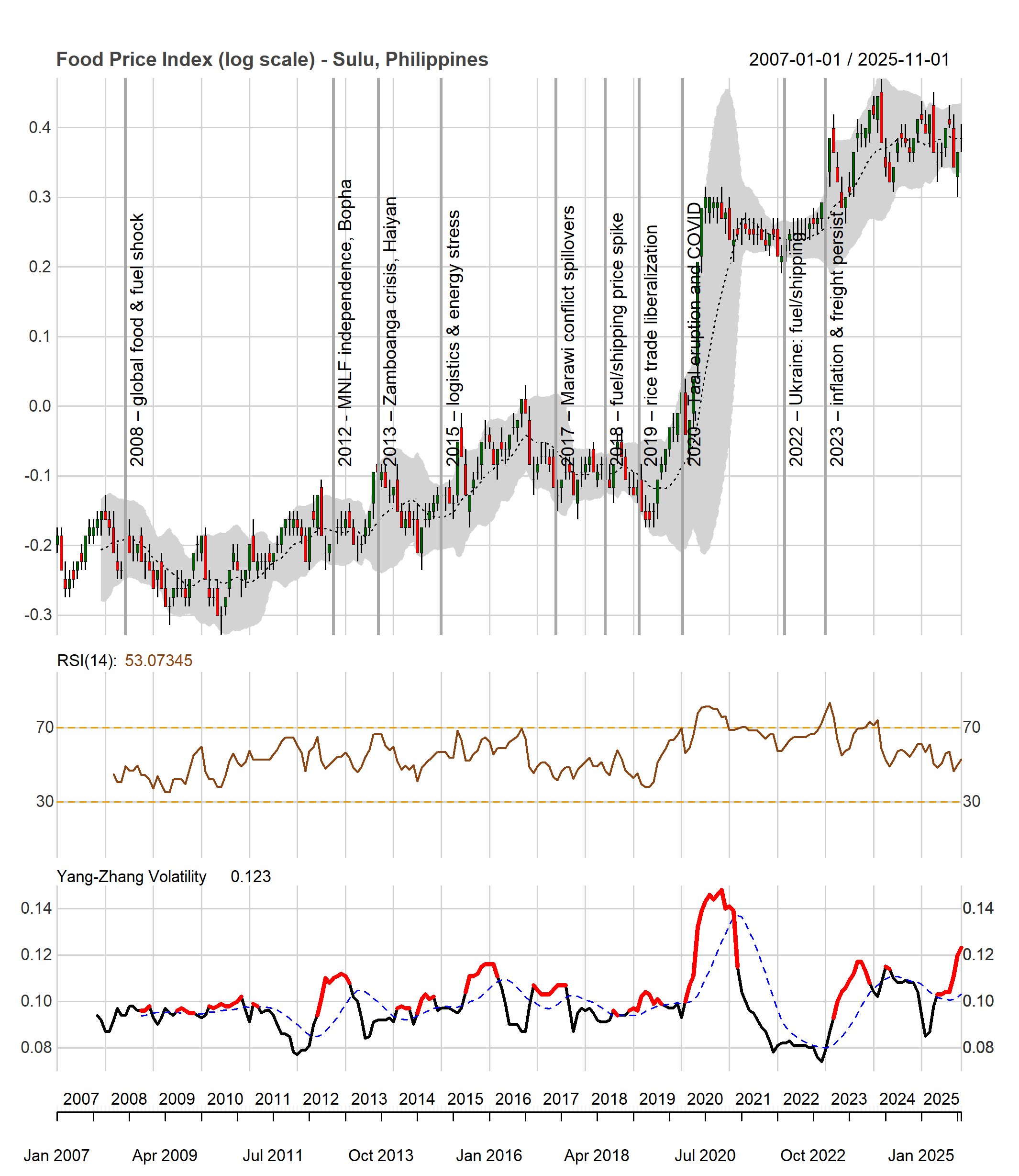}
    \caption{Philippines (Sulu): technical setup and detected stress episodes. Top panel: log food prices. Middle panel: RSI with 30 and 70 reference levels. Bottom panel: Yang--Zhang annualized volatility, its 12-month moving average, and detected high-volatility episodes (red segments).}
    \label{fig:phl_signal}
\end{figure}

The Philippines provides a complementary test of portability beyond settings defined by extremes. We focus on Sulu in the Zamboanga Peninsula, a relatively remote market shaped by localized insecurity and logistics frictions, and policy shifts affecting exposure to global staple-market conditions. 

Flagged volatility segments highlight recurring stress episodes consistent with intermittent access and discontinuous price formation. In the early part of the series, the 2007--08 global food and fuel crisis does not appear as a dominant contiguous episode, suggesting muted transmission into observed dispersion relative to local dynamics. Instead, pre-2019 volatility clusters are more closely associated with domestic disruption and logistics stress, including episodes in the early-to-mid 2010s and again in 2015--2018.

A key policy discontinuity occurs in 2019 with the Rice Tariffication Law, which replaced quantitative import restrictions with a tariff-based regime and altered import incentives and price expectations. In the Sulu series this shift is not marked by a single defining spike, but it precedes a period in which externally driven disruptions become more apparent. The clearest example is COVID-19 in 2020, which coincides with the largest volatility spike in the sample, consistent with abrupt mobility and supply-chain disruption. RSI rises during the COVID-19 period but remains within the neutral band, whereas volatility clearly flags the episode as exceptional. The 2022 Russia--Ukraine shock does not register as a sharp standalone episode, although the post-2019 period exhibits more frequent clustering alongside sustained freight and inflation pressures that followed. While these labels remain \textit{ex post} and partly judgment-based---especially when shocks overlap---the overall pattern supports the operational point that volatility remains informative even when policy change reshapes a market's exposure to national and global conditions.

\section{Discussion}\label{sec:discussion}

\subsection{Cross-country synthesis}

Table~\ref{tab:summary} summarizes the main shock categories that coincide with flagged volatility episodes across the five applications. The mapping is necessarily approximate: event labels are used for \textit{ex post} interpretation, and attribution depends on the granularity of available documentation and on analyst judgment when multiple drivers overlap. The objective is not to build a definitive classification, but to illustrate that the volatility workflow consistently highlights a broad and policy-relevant range of disruptions across diverse contexts.

\begin{table}[htbp]
\centering
\caption{Summary of detected volatility episodes by shock type across five countries (based on flagged Yang--Zhang volatility segments in red).}
\label{tab:summary}
\small
\begin{tabular}{lccccc}
\hline
Shock type & Sudan & Somalia & Cameroon & Haiti & Philippines \\
\hline
Global food/fuel crisis (2007--08) & Yes & Yes & --  & Yes & -- \\
Global shock (2022 Ukraine)        & --  & --  & Yes & --  & -- \\
Armed conflict / insecurity        & Yes & Yes & Yes & Yes & Yes \\
Natural disaster (flood/drought)   & --  & Yes & Yes & Yes & Yes \\
Natural disaster (earthquake/volcano) & -- & -- & -- & Yes & Yes \\
Policy shock (subsidy/trade/regime shift) & Yes & -- & Yes & Yes & Yes \\
Political transition / crisis      & Yes & --  & --  & Yes & -- \\
COVID-19                           & --  & Yes & --  & Yes & Yes \\
\hline
\end{tabular}
\end{table}

Several patterns emerge. First, volatility responds to global-to-local transmission when the shock is large and the market is exposed. The 2007--08 food and fuel crisis appears as a distinct volatility episode in Sudan, Somalia, and Haiti, consistent with uneven pass-through of international costs into domestic staple prices in import- and fuel-exposed systems \citep{fao2009,hlpe2011,gilbert2010}. By contrast, the 2022 Russia--Ukraine shock is most clearly detected in Cameroon in the plotted series, suggesting stronger transmission through energy and trade channels in that setting, while the other cases exhibit weaker or less distinct volatility responses at monthly frequency.

Second, locally generated disruptions are detected in ways that cannot be inferred from global benchmarks alone. Conflict and local insecurity coincide with flagged volatility episodes in all five settings, consistent with fragmentation of market access, higher corridor risk, and intermittency of supply. Natural hazards generate identifiable volatility regimes, including drought and flood stress (Somalia, Cameroon, Haiti, and the Philippines) and large discrete disruptions such as Haiti's earthquake and the Philippines' volcanic event. These patterns support volatility measures acting as a reduced-form indicator of impaired market integration and unstable price formation.

Third, the largest and most persistent volatility regimes tend to arise under compound stress, when multiple channels overlap or reinforce one another. Somalia's famine period combines climatic stress with fragile access and market constraints. Sudan's 2023--2025 episode reflects war-driven market dysfunction interacting with siege dynamics and macroeconomic instability. Haiti's 2019--2024 period illustrates how repeated political disruption, fuel access constraints, and insecurity can jointly sustain elevated volatility in an urban import hub. In such settings, volatility serves as a compact summary measure of cumulative market stress rather than a one-to-one proxy for a single event.

Finally, the cross-country cases clarify why dispersion-based monitoring can complement momentum-type indicators. RSI and related measures primarily respond to persistent directional movements, whereas volatility rises under both price surges and abrupt reversals, including episodes dominated by intermittency and access constraints rather than smooth trend inflation. Conceptually, this parallels microstructure settings in which frictions and widening spreads distort price discovery and increase short-horizon dispersion even when long-run price-level adjustment is ambiguous \citep{hasbrouck2002,corwinschultz2012}. Across the five cases, volatility flags a meaningful share of documented stress episodes during which RSI remains within neutral bounds (30--70), including the 2020 COVID-19 disruption in Sulu, the 2023--2024 floods in Somalia, and the post-2022 cost-transmission regime in Cameroon's Far North.

\subsection{Implications for market monitoring}

The cross-country results highlight several advantages of incorporating volatility measures into market monitoring systems. First, volatility is symmetrically sensitive to shocks. Unlike trend-based indicators such as inflation, RSI, or MACD \citep{appel1979,wilder1978}, which primarily flag persistent directional movements, volatility responds to both price increases and decreases. This feature is particularly valuable in environments where supply and demand disruptions partially offset each other, for example when insecurity reduces production and market access while also compressing purchasing power, or when aid inflows temporarily suppress prices before sharp increases re-emerge as supplies tighten.

Second, volatility provides a compact summary of market stress. By condensing the joint effects of macro instability, policy changes, and local disruptions into a single scalar indicator, it offers a practical signal for dashboards and routine surveillance. In operational settings, classifying volatility relative to its historical distribution can support rapid triage of abnormal conditions and can complement other signals such as price levels, exchange rate movements, or conflict and weather information. Conceptually, sustained increases in dispersion are also consistent with a shift toward weaker market liquidity and higher trading frictions, in the sense that prices adjust more discontinuously when transaction costs rise and market participation becomes constrained \citep{adrianetal2017,schwarz2018}.

Third, the estimators are computationally simple and stable. Range-based measures can be computed directly once a period's OHLC values are observed, without the repeated re-estimation required by GARCH-type models and without backward revisions to previously reported volatility series. This stability makes them well-suited for automated monitoring, pre-defined alert thresholds, and consistent reporting across markets and time.

Fourth, the approach generalizes across contexts. The five case studies span different regions, income levels, market structures, and shock types, yet the volatility measures produce consistent signals of elevated stress around documented events. This pattern suggests that the estimators capture fundamental properties of market dysfunction under stress, rather than reflecting context-specific features of a single country or commodity.

\subsection{Limitations}

Several limitations warrant emphasis. First, the RTP OHLC series are model-based rather than directly observed. While the fiGARCH--GED framework is designed to capture realistic within-period dynamics \citep{baillie1996,andree2021_inflation,andreepape2023}, the quality of volatility estimates ultimately depends on the underlying model, the coverage and reliability of input data, and the imputation process. Section~\ref{sec:OHLC_relation} shows that range-based estimators remain informative as long as the OHLC structure encodes meaningful within-period dispersion; when this structure collapses (e.g., when conditional variance vanishes or when high and low prices converge to the close), the estimators reduce to close-to-close measures and the incremental information from ranges disappears. The empirical results indicate that this collapse does not occur in the settings examined, but direct validation of the synthetic OHLC series against observed intra-period data, where available, would further strengthen confidence in the approach.

Second, the detection rule relies on a simple threshold defined relative to the historical distribution and recent moving averages. More adaptive approaches, including time-varying thresholds, supervised classifiers, or integration with additional covariates, may improve precision and reduce false alarms. The advantage of the current rule is transparency and ease of implementation in operational systems.

Third, volatility remains a reduced-form indicator. While the case studies show close alignment with documented shocks, elevated volatility does not identify causal mechanisms, and similar signatures may arise from different sources such as insecurity, climate shocks, macroeconomic instability, or policy interventions. Combining volatility with structural models or complementary covariates could help distinguish mechanisms and quantify the incremental contribution of volatility beyond mean price changes.

Fourth, the analysis focuses on market-level outcomes and does not trace transmission to household welfare. The literature documents strong links between food prices, nutrition, poverty, and social stability \citep{bellemare2015,headeyruel2023,amolegbe2021}, but the welfare consequences of volatility specifically remain less well characterized. Linking volatility indicators to household microdata and outcome measures remains an important direction for future work, particularly for calibrating thresholds used in operational triggers.

Finally, while the five case studies provide diverse illustration, broader evaluation across the full set of markets covered by RTP would allow more systematic assessment of detection accuracy, false positive rates, and the conditions under which volatility indicators perform best. Such evaluation would benefit from integration with independent event datasets and monitoring frameworks used in practice.

\section{Conclusion}\label{sec:conclusion}

This paper demonstrates that range-based volatility estimators from financial econometrics can be productively applied to monitor market stress across diverse development contexts. Using OHLC price data from the World Bank's Real-Time Prices system, we show that the Yang--Zhang volatility estimator consistently captures responses to documented shocks---including global commodity crises, armed conflict, natural disasters, and policy changes---across five country case studies spanning Sub-Saharan Africa, the Middle East, the Caribbean, and Southeast Asia.

The results highlight volatility as a complement to price-level monitoring. Traditional momentum indicators such as the RSI can produce muted or misleading signals during crises where supply and demand disruptions partially offset directional price movements. By contrast, volatility captures market dysfunction that manifests in widening intra-period price dispersion, abrupt reversals, and intermittent price formation rather than sustained trends, providing information that price levels alone cannot reveal.

Beyond food markets, the approach generalizes in principle to any setting where OHLC price series are available. In financial risk management, volatility indicators can support portfolio monitoring, hedging decisions, and stress detection in commodity-linked exposures. In trade and supply-chain analysis, elevated volatility can provide early signals of disruptions related to border frictions, logistics constraints, or policy uncertainty, especially when official statistics are delayed or insufficiently granular. In macroeconomic surveillance, exchange-rate and energy-price volatility can serve as compact indicators of broader instability. In humanitarian and development operations, volatility can complement price levels and momentum measures in early-warning systems and preparedness frameworks by helping distinguish inflationary pressure from market dislocation. Finally, volatility measures can support policy evaluation by providing a transparent metric to assess whether interventions such as subsidy reforms, trade liberalization, or price controls dampen or amplify market instability.

More broadly, the methodology illustrates how techniques developed for financial markets can be translated to operational monitoring. Range-based volatility estimators are computationally simple, require no model re-estimation, and yield stable historical series that can be deployed as automated alerts against pre-defined thresholds. As real-time data systems expand and automated analytics mature, compact and interpretable indicators like volatility can help convert high-frequency price information into actionable signals. When embedded in monitoring dashboards and linked to response protocols, such measures can support faster recognition of emerging market stress and more timely, targeted policy and operational responses.

%\section*{Acknowledgments}

%The author gratefully acknowledges colleagues in the World Bank Data Group and partners involved in the development of the Real-Time Prices system and the Joint Monitoring Report as well as the Sudan emergency response team. Any remaining errors are the author's own.

%\subsection*{Funding}

%This work was supported by Food Systems 2030 (TF0C7822) and the Umbrella Facility for Trade Multi-Donor Trust Fund 2.0 (TF074184). It was carried out as part of the author's employment at the World Bank.

%\subsection*{Conflicts of Interest}

%The author declares that there is no conflict of interest regarding the publication of this article.

\subsection*{Data and Code Availability}

The RTP-based price series used in this paper are available through the World Bank's data portals and microdata library, subject to the usual terms of use. All estimators are implemented in \textsf{R}. Code and data to reproduce results and figures are archived on Zenodo at \url{https://doi.org/10.5281/zenodo.18846560}.

\bibliographystyle{apalike}
\bibliography{sample}

\appendix

\section{Additional Figures}\label{app:additional_figures}

\begin{figure}[htbp]
    \centering
    \includegraphics[width=0.90\textwidth]{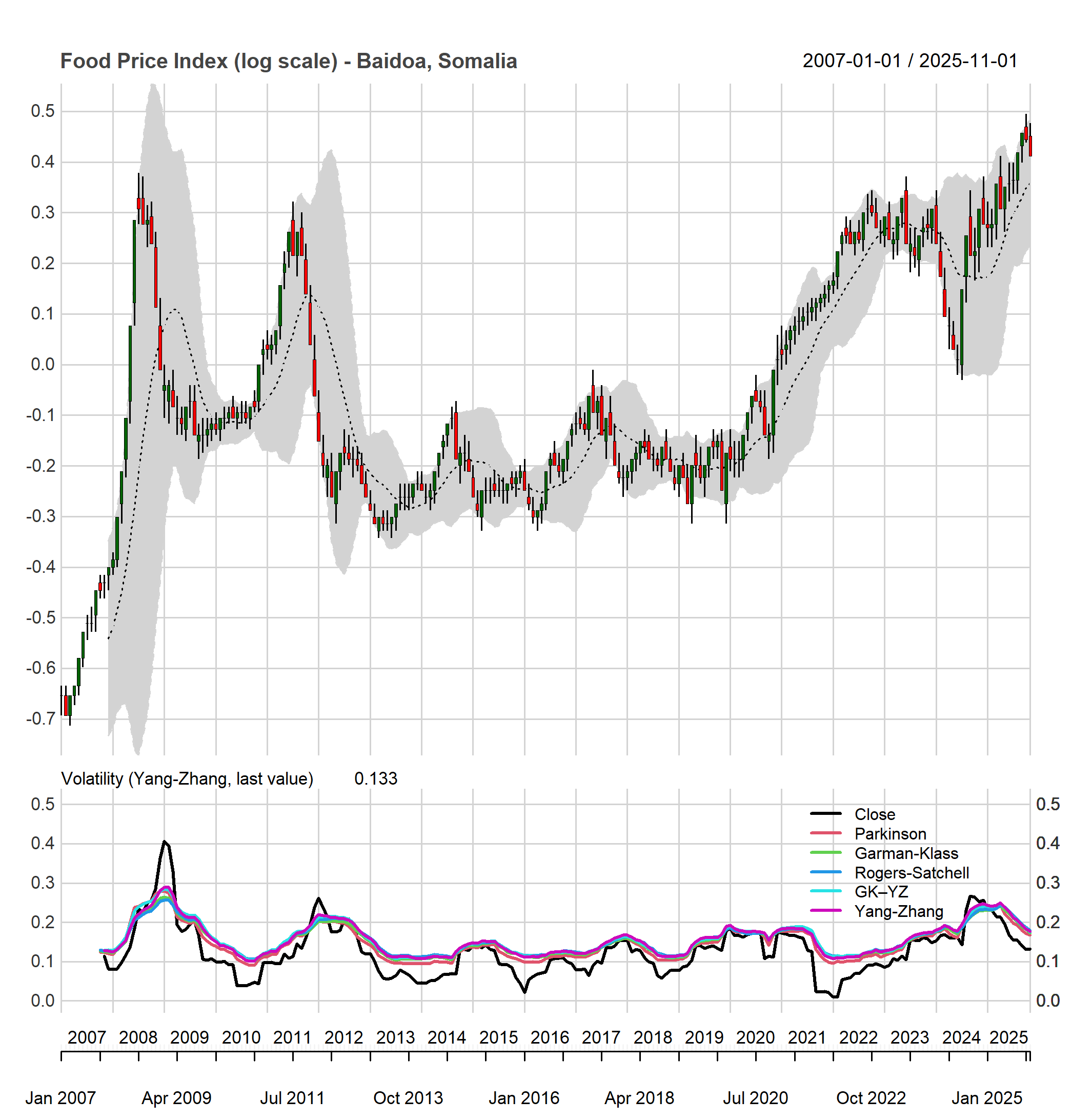}
    \caption{Open, High, Low, Close food prices (index, log) in Baidoa, Somalia, plotted on a candlestick chart together with six OHLC-based volatility metrics.}
    \label{fig:vol_all_som}
\end{figure}

\begin{figure}[htbp]
    \centering
    \includegraphics[width=0.90\textwidth]{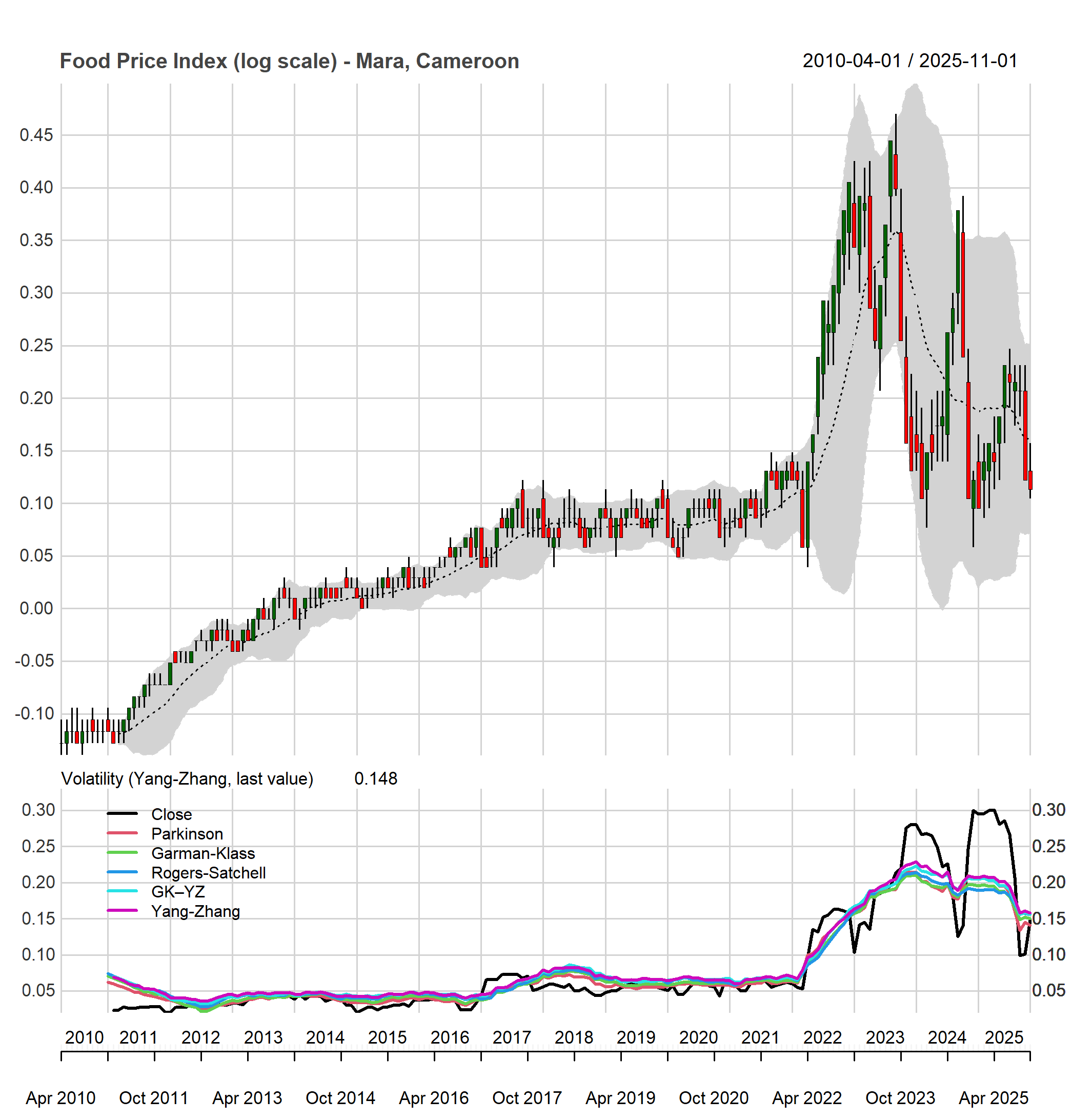}
    \caption{Open, High, Low, Close food prices (index, log) in Far North, Cameroon, plotted on a candlestick chart together with six OHLC-based volatility metrics.}
    \label{fig:vol_all_cmr}
\end{figure}

\begin{figure}[htbp]
    \centering
    \includegraphics[width=0.90\textwidth]{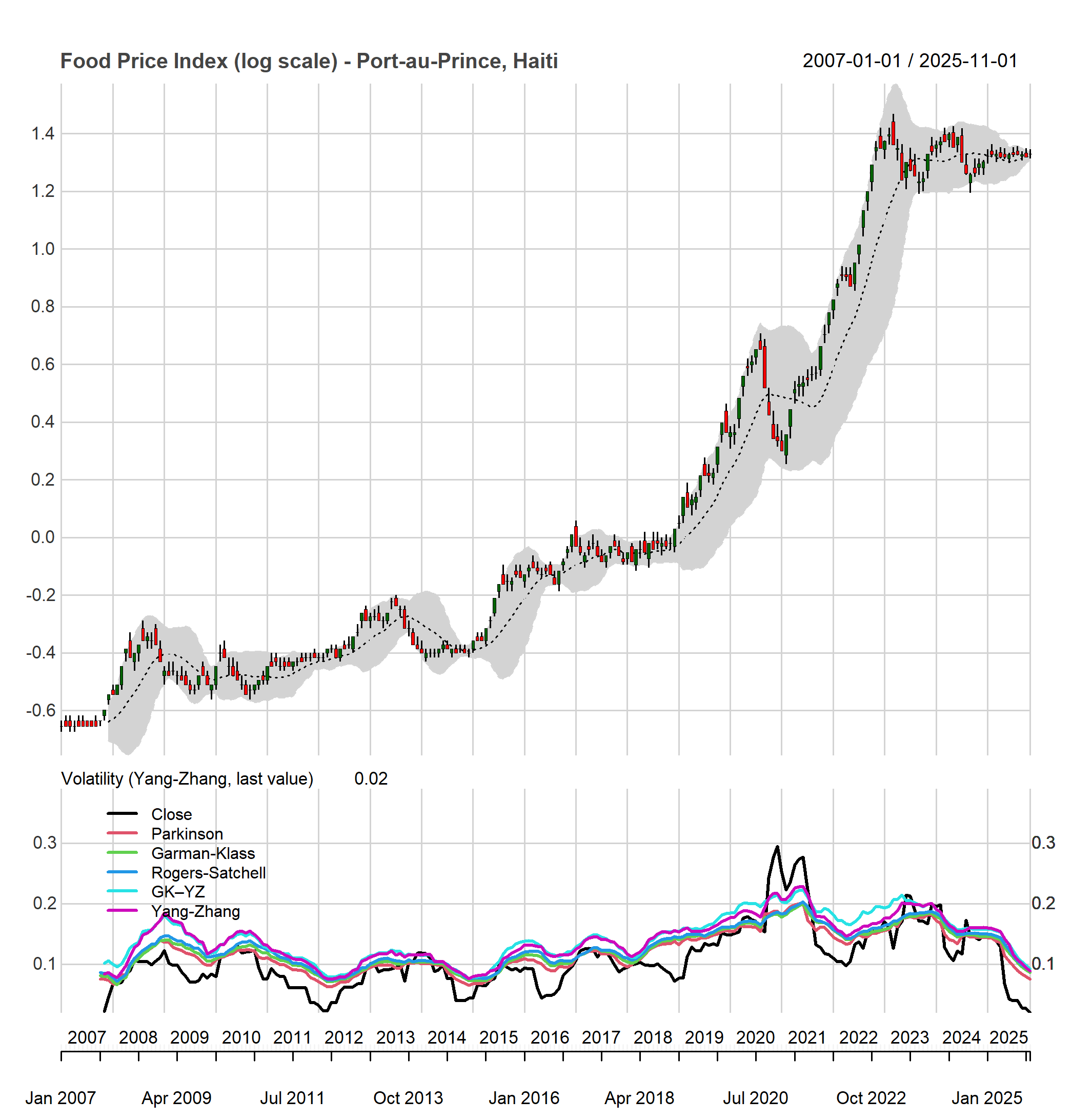}
    \caption{Open, High, Low, Close food prices (index, log) in Port-au-Prince, Haiti, plotted on a candlestick chart together with six OHLC-based volatility metrics.}
    \label{fig:vol_all_hti}
\end{figure}

\begin{figure}[htbp]
    \centering
    \includegraphics[width=0.90\textwidth]{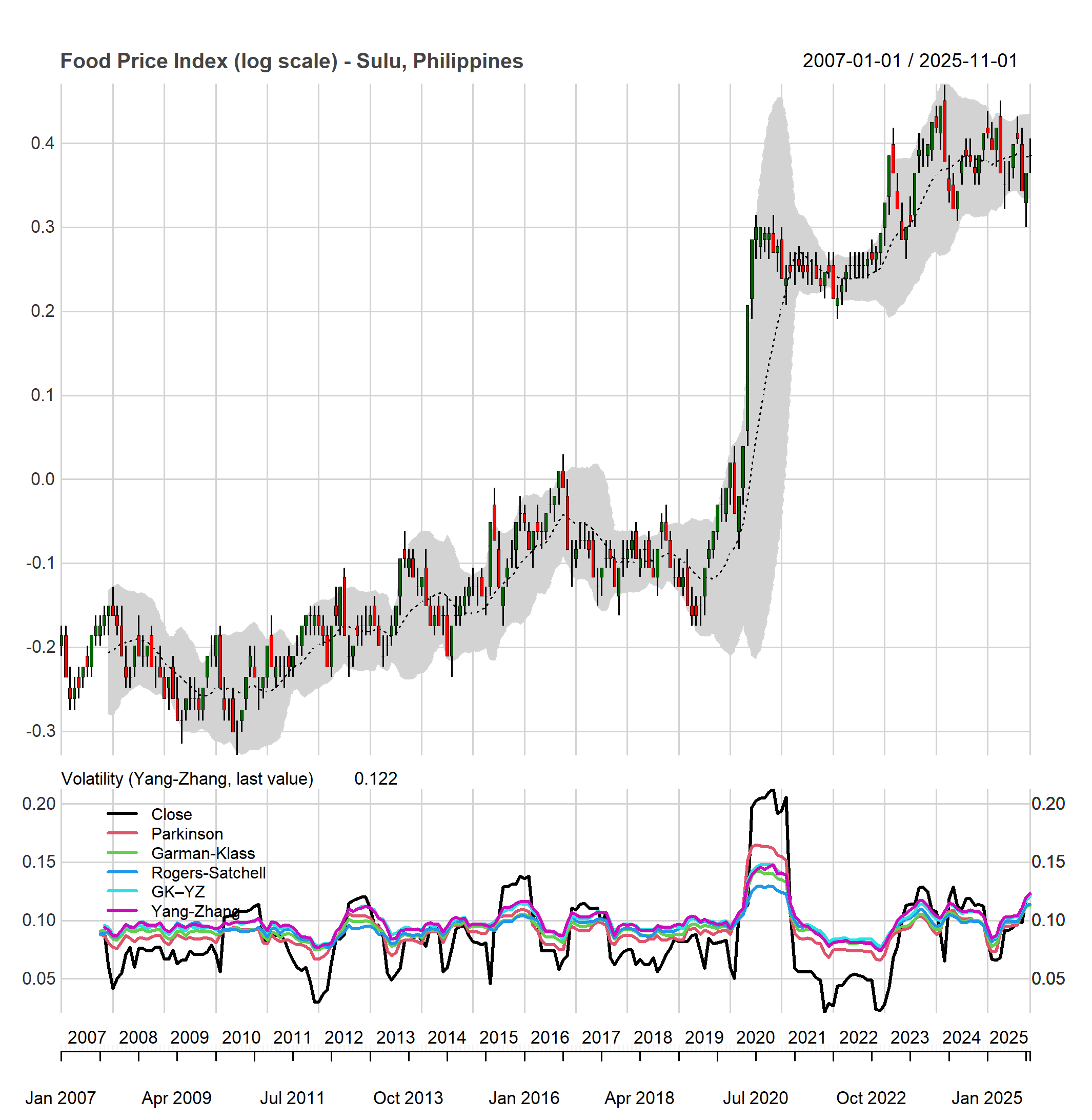}
    \caption{Open, High, Low, Close food prices (index, log) in Sulu, Philippines, plotted on a candlestick chart together with six OHLC-based volatility metrics.}
    \label{fig:vol_all_phl}
\end{figure}

This appendix provides additional figures showing the different volatility indicators in Somalia (Figure~\ref{fig:vol_all_som}), Cameroon (Figure~\ref{fig:vol_all_cmr}), Haiti (Figure~\ref{fig:vol_all_hti}), and the Philippines (Figure~\ref{fig:vol_all_phl}).

\section{Supplementary Materials: Sudan Case Study (Al Fashir)}\label{app:sudan_events}

This appendix provides an annotated timeline of major national and Darfur-specific episodes used to interpret volatility dynamics in Al Fashir. The timeline is compiled from human rights reporting, humanitarian monitoring systems, governance and political assessments, and international media coverage. The events are used solely for \textit{ex post} interpretation of detected volatility segments and are not employed for causal identification.

\begingroup
\footnotesize
\setlength{\tabcolsep}{3pt}
\renewcommand{\arraystretch}{1.07}
\setlength{\LTleft}{0pt}
\setlength{\LTright}{0pt}
\sloppy

\begin{longtable}{@{}P{1.55cm} P{4.10cm} P{1.55cm} P{6.05cm} S{2.05cm}@{}}
\caption{Al Fashir (Sudan): annotated event timeline for interpreting detected volatility episodes.}\label{tab:sudan_timeline_full}\\
\hline
\textbf{Period} &
\textbf{Episode and context} &
\textbf{Scale} &
\textbf{Expected market mechanism and interpretation} &
\textbf{Key sources} \\
\hline
\endfirsthead

\hline
\textbf{Period} &
\textbf{Episode and context} &
\textbf{Scale} &
\textbf{Expected market mechanism and interpretation} &
\textbf{Key sources} \\
\hline
\endhead

\hline
\multicolumn{5}{r}{\textit{Continued on next page}}\\
\endfoot

\hline
\endlastfoot

2007--2008 &
Global food and fuel price crisis transmitted to low-income import-dependent economies. &
Global &
A global import-cost shock raises staple price levels and increases within-period dispersion as pass-through varies over time and across supply channels. This can generate sustained high dispersion even when the overall price trend remains upward. &
\citep{headey2010} \\

2007--2009 &
Continued Darfur conflict dynamics and displacement pressures affecting market access around Al Fashir. &
Darfur &
Insecurity, displacement, and disrupted trade routes amplify intermittency in local supplies and raise the frequency of abrupt price movements. In volatility terms, this contributes to sustained abnormal dispersion consistent with localized market fragmentation. &
\citep{amnesty2014} \\

2010 (Apr.) &
General elections under fragile national conditions and persistent insecurity in Darfur. &
National \& Darfur &
Political uncertainty and localized insecurity can increase risk premia in trading corridors, disrupt market integration, and raise the variance of price changes even absent a discrete national macro shock. &
\citep{hassan2019} \\

2011--2014 &
South Sudan secession and fiscal rupture followed by austerity measures, subsidy cuts, and protest waves. &
National &
Macroeconomic adjustment and policy instability can produce sharp price jumps and partial reversals, increasing within-month dispersion through sudden supply interruptions and uncertainty about policy continuity. &
\citep{hrw2013} \\

2016 &
Jebel Marra offensive and renewed violence in Darfur. &
Darfur &
Conflict intensification can disrupt rural production and access to market corridors, reducing the regularity of staple inflows and increasing intermittency in availability. &
\citep{amnesty2016} \\

2017--2019 &
Deepening macroeconomic crisis culminating in bread-price protests and the fall of President Bashir (Dec.\ 2018--Apr.\ 2019). &
National &
A transition from chronic inflation to an acute crisis episode: shortages, depreciation, and political instability cause repeated supply breaks and abrupt corrections, increasing dispersion. &
\citep{hassan2019} \\

2019--2022 &
Transitional reforms under severe macro stress, followed by the October 2021 coup and renewed political uncertainty. &
National &
Weak policy credibility and institutional instability can sustain a high-volatility plateau through repeated repricing, shifting expectations, and enforcement discontinuities. &
\citep{hassan2019} \\

2022--mid-2023 &
Post-coup stagnation and mounting food-security stress preceding the outbreak of full-scale war. &
National \& Darfur &
Chronic inflation and deteriorating terms of trade can cause volatility to drift upward, appearing as clusters of high-volatility flags consistent with incremental erosion of market functioning. &
\citep{fewsnet2024} \\

mid-2023--2025 &
SAF--RSF war and extended siege dynamics affecting Al Fashir; severe food-security outcomes and famine conditions. &
National \& Darfur (Al Fashir focus) &
Physical insecurity and blockade-like conditions generate extreme within-month dispersion, sharp jumps, and partial corrections, consistent with structural market failure and fragmented price formation. &
\citep{fewsnet2024} \\

\end{longtable}
\endgroup

Table~\ref{tab:sudan_mechanisms} summarizes the dominant interpretation channels associated with each event class. The goal is to standardize narrative interpretation across cases while keeping the framework descriptive rather than causal.

\begin{table}[htbp]
\centering
\begingroup
\footnotesize
\setlength{\tabcolsep}{3pt}
\renewcommand{\arraystretch}{1.07}
\sloppy

\begin{tabularx}{\textwidth}{@{}P{3.0cm} Y P{3.1cm}@{}}
\hline
\textbf{Event class} & \textbf{Typical volatility mechanism in market price data} & \textbf{Examples in Sudan} \\
\hline
Global price shock &
Pass-through of international price increases can be uneven across months due to import timing, credit constraints, and heterogeneous local supply substitution. This often produces elevated dispersion beyond what level-based trend measures capture. &
2007--08 global food price crisis \\
\hline
Macroeconomic rupture / policy reform &
Currency depreciation, subsidy removal, and credibility shocks can generate repeated step changes and reversals in local prices as expectations update and policy implementation fluctuates. Dispersion increases even when the average trend remains upward. &
Post-2011 adjustment; reforms under transition (2019--2022) \\
\hline
Conflict escalation / corridor disruption &
Insecurity and violence disrupt flows, increase transaction costs, and fragment markets. Prices become intermittent and discontinuous, often yielding sharp volatility spikes even when the direction of price changes is unclear. &
Jebel Marra (2016); war and siege dynamics (2023--2025) \\
\hline
Political upheaval / institutional shocks &
Transitions, coups, and governance breakdown can operate as uncertainty shocks that amplify dispersion through behavioral responses (hoarding, speculation, enforcement discontinuities) rather than through a single supply or demand lever. &
2010 elections; October 2021 coup \\
\hline
Chronic stress / deterioration &
A gradual worsening of conditions may show up as clusters of smaller high-volatility flags rather than one discrete event. This pattern can matter operationally because it signals sustained fragility even without a headline trigger. &
2022--mid-2023 pre-war stress \\
\hline
\end{tabularx}

\caption{Interpretation guidance for relating contextual events to detected volatility segments in Al Fashir.}
\label{tab:sudan_mechanisms}
\endgroup
\end{table}

\section{Supplementary Materials: Somalia Case Study (Baidoa)}\label{app:somalia_events}

This appendix provides an annotated timeline of major national and subnational episodes used to interpret volatility dynamics in Baidoa. The timeline is compiled from humanitarian and food-security monitoring products, UN reporting, and public documentation. The events are used solely for \textit{ex post} interpretation of detected volatility segments and are not employed for causal identification.

\begingroup
\footnotesize
\setlength{\tabcolsep}{3pt}
\renewcommand{\arraystretch}{1.07}
\setlength{\LTleft}{0pt}
\setlength{\LTright}{0pt}
\sloppy

\begin{longtable}{@{}P{1.55cm} P{4.10cm} P{1.55cm} P{6.05cm} S{2.05cm}@{}}
\caption{Baidoa (Somalia): annotated event timeline for interpreting detected volatility episodes.}\label{tab:somalia_timeline_full}\\
\hline
\textbf{Period} &
\textbf{Episode and context} &
\textbf{Scale} &
\textbf{Expected market mechanism and interpretation} &
\textbf{Key sources} \\
\hline
\endfirsthead

\hline
\textbf{Period} &
\textbf{Episode and context} &
\textbf{Scale} &
\textbf{Expected market mechanism and interpretation} &
\textbf{Key sources} \\
\hline
\endhead

\hline
\multicolumn{5}{r}{\textit{Continued on next page}}\\
\endfoot

\hline
\endlastfoot

2007--2008 &
Global food price crisis transmitted to Somalia during a period of structural vulnerability. &
Global \& Somalia &
A global import-cost shock raises staple price levels and increases within-period dispersion as pass-through and availability vary over time. &
\citep{headey2010} \\

2011 &
Severe food-security crisis and famine in parts of southern Somalia. &
National \& South-central &
A compound shock environment contributes to sharp price dislocations and abnormal dispersion consistent with intermittent supply and breakdowns in market clearing. &
\citep{maxwell2012} \\

2014--2015 &
Renewed stress associated with weak seasonal performance and disruptions affecting Bay/Bakool livelihoods and trade flows. &
Subnational (Bay/Bakool) &
Localized constraints reduce the regularity of inflows and raise dispersion through intermittent supply and shifting expectations. &
\citep{checchi2023} \\

2016--2017 &
Drought emergency and warnings of renewed famine risk; large-scale impacts mitigated by early response. &
National \& South-central &
Drought generates elevated dispersion through production shortfalls, income losses, and shifting trade patterns, typically appearing as sustained high volatility. &
\citep{checchi2023} \\

2020 &
COVID-19 related disruptions coinciding with desert locust impacts. &
National \& regional &
Simultaneous shocks affect logistics and production risk, amplifying within-month dispersion even when average price levels move gradually. &
\citep{workie2020} \\

2022 &
Global price transmission following the Russia--Ukraine war. &
Global \& Somalia &
Externally driven cost increases raise dispersion through uneven pass-through and intermittent access conditions. &
\citep{arab2022} \\

2023--2024 &
Severe flooding episodes affecting large parts of Somalia. &
National \& regional &
Flooding disrupts transport and market functioning, producing sharp volatility spikes reflecting localized shortages and corrections as flows resume. &
\citep{ocha2024_somalia_floods} \\

\end{longtable}
\endgroup

Table~\ref{tab:somalia_mechanisms} summarizes the dominant interpretation channels associated with each event class. The goal is to standardize narrative interpretation across cases while keeping the framework descriptive rather than causal.

\begin{table}[htbp]
\centering
\begingroup
\footnotesize
\setlength{\tabcolsep}{3pt}
\renewcommand{\arraystretch}{1.07}
\sloppy

\begin{tabularx}{\textwidth}{@{}P{3.0cm} Y P{3.1cm}@{}}
\hline
\textbf{Event class} & \textbf{Typical volatility mechanism in market price data} & \textbf{Examples in Somalia} \\
\hline
Global price shock &
Uneven pass-through of international food and fuel prices increases dispersion through import timing, exchange-rate pressures, credit frictions, and heterogeneous substitution in local markets. &
2007--08 crisis; 2022 global transmission \\
\hline
Drought / rainfall failure &
Production shortfalls and income losses raise dispersion through intermittent availability, shifting trade flows, and nonlinear responses in purchasing power and demand compression. &
2011 famine context; 2016--2017 drought \\
\hline
Flood / transport disruption &
Infrastructure damage and corridor disruption create localized shortages and sharp corrections as access fluctuates over time. &
2023--2024 floods \\
\hline
Compound disruption (health, pests, logistics) &
Simultaneous shocks can impair both supply and logistics while raising uncertainty, generating volatility even when levels move gradually. &
COVID-19 and locust upsurge (2020) \\
\hline
Localized livelihood and access stress &
Subnational disruptions can appear as clusters of high-volatility flags driven by intermittency in flows to key markets and relief hubs. &
Bay/Bakool stress (2014--2015) \\
\hline
\end{tabularx}

\caption{Interpretation guidance for relating contextual events to detected volatility segments in Baidoa.}
\label{tab:somalia_mechanisms}
\endgroup
\end{table}

\section{Supplementary Materials: Cameroon Case Study (Far North)}\label{app:cameroon_events}

This appendix provides an annotated timeline of major episodes used to interpret volatility dynamics in Cameroon's Far North. The timeline is compiled from conflict reporting and humanitarian monitoring, flood and emergency assessments, policy documentation, and public sources. These materials are used solely for \textit{ex post} interpretation of detected volatility segments and are not employed for causal identification.

\begingroup
\footnotesize
\setlength{\tabcolsep}{3pt}
\renewcommand{\arraystretch}{1.07}
\setlength{\LTleft}{0pt}
\setlength{\LTright}{0pt}
\sloppy

\begin{longtable}{@{}P{1.55cm} P{4.10cm} P{1.55cm} P{6.05cm} S{2.05cm}@{}}
\caption{Cameroon (Far North): annotated event timeline for interpreting detected volatility episodes.}\label{tab:cameroon_timeline_full}\\
\hline
\textbf{Period} &
\textbf{Episode and context} &
\textbf{Scale} &
\textbf{Expected market mechanism and interpretation} &
\textbf{Key sources} \\
\hline
\endfirsthead

\hline
\textbf{Period} &
\textbf{Episode and context} &
\textbf{Scale} &
\textbf{Expected market mechanism and interpretation} &
\textbf{Key sources} \\
\hline
\endhead

\hline
\multicolumn{5}{r}{\textit{Continued on next page}}\\
\endfoot

\hline
\endlastfoot

2014--2016 &
Intensification of insecurity and insurgency spillovers in the Far North. &
Subnational \& cross-border &
Insecurity increases transaction costs and intermittency of supply, contributing to higher dispersion through market fragmentation and discontinuous trading. &
\citep{icg2018_cameroon_far_north} \\

2017 &
Flood-related shocks affecting livelihoods and transport access. &
Subnational &
Flooding disrupts production and infrastructure, creating localized shortages and unstable access that elevate dispersion. &
\citep{unicef2017_cameroon_floods} \\

2019--2020 &
Cross-border trade constraints associated with Nigeria's border policy episode. &
Regional \& cross-border &
Border frictions reduce arbitrage and raise dispersion through irregular availability and uneven pass-through. &
\citep{aremu2023} \\

2020 &
COVID-19 mobility restrictions and logistics disruptions. &
National &
Movement constraints disrupt supply chains and raise dispersion through irregular inflows and episodic repricing. &
\citep{amare2020_covid_food_security_africa} \\

2022 &
Flood emergencies and renewed climate-related disruption. &
Subnational &
Transport impairment and localized production losses increase intermittency in supplies and raise volatility. &
\citep{ocha2022_cameroon_floods} \\

2022 &
Global commodity-price and energy-cost pressures following the Russia--Ukraine shock. &
Global \& Cameroon &
Externally driven increases in food and fuel costs transmit through import and transport channels, elevating dispersion through uneven pass-through. &
\citep{arab2022} \\

2023 &
Fuel price adjustment and cost shock to transport and distribution margins. &
National &
Discrete increases in energy costs generate abrupt repricing and elevated dispersion, especially in remote areas where transport margins are large. &
\citep{imf2023_cameroon} \\

2024 &
Renewed flood impacts and humanitarian access constraints. &
Subnational &
Flooding and access limitations sustain intermittency in supply conditions and localized market fragmentation. &
\citep{ocha2024_cmr} \\

\end{longtable}
\endgroup

Table~\ref{tab:cameroon_mechanisms} summarizes the dominant interpretation channels associated with each event class. The goal is to standardize narrative interpretation across cases while keeping the framework descriptive rather than causal.

\begin{table}[htbp]
\centering
\begingroup
\footnotesize
\setlength{\tabcolsep}{3pt}
\renewcommand{\arraystretch}{1.07}
\sloppy

\begin{tabularx}{\textwidth}{@{}P{3.0cm} Y P{3.1cm}@{}}
\hline
\textbf{Event class} & \textbf{Typical volatility mechanism in market price data} & \textbf{Examples in Cameroon} \\
\hline
Conflict / insecurity &
Higher corridor risk and transaction costs fragment markets and generate intermittent supply, producing abrupt jumps and partial corrections. &
Insurgency spillovers (mid-2010s) \\
\hline
Flood / transport disruption &
Infrastructure damage and access constraints create localized shortages and unstable inflows, often appearing as clustered volatility flags. &
Flood episodes (2017; 2022; 2024) \\
\hline
Cross-border policy disruption &
Border frictions weaken arbitrage and raise dispersion through irregular availability and uneven pass-through across locations and months. &
Nigeria border policy episode (2019--2020) \\
\hline
Mobility and logistics shock &
Restrictions and uncertainty disrupt supply chains and distribution margins even if price levels move gradually. &
COVID-19 period (2020) \\
\hline
Energy / cost transmission &
Fuel and freight shocks transmit quickly through distribution margins, generating discrete repricing and elevated dispersion. &
Fuel price adjustment (2023); global cost pressures (2022) \\
\hline
\end{tabularx}

\caption{Interpretation guidance for relating contextual events to detected volatility segments in Cameroon's Far North.}
\label{tab:cameroon_mechanisms}
\endgroup
\end{table}

\section{Supplementary Materials: Haiti Case Study (Port-au-Prince)}\label{app:haiti_events}

This appendix provides an annotated timeline of major episodes used to interpret volatility dynamics in Port-au-Prince. The timeline is compiled from earthquake and disaster documentation, humanitarian monitoring, governance and political reporting, and public sources. These materials are used solely for \textit{ex post} interpretation of detected volatility segments and are not employed for causal identification.

\begingroup
\footnotesize
\setlength{\tabcolsep}{3pt}
\renewcommand{\arraystretch}{1.07}
\setlength{\LTleft}{0pt}
\setlength{\LTright}{0pt}
\sloppy

\begin{longtable}{@{}P{1.55cm} P{4.10cm} P{1.55cm} P{6.05cm} S{2.05cm}@{}}
\caption{Haiti (Port-au-Prince): annotated event timeline for interpreting detected volatility episodes.}\label{tab:haiti_timeline_full}\\
\hline
\textbf{Period} &
\textbf{Episode and context} &
\textbf{Scale} &
\textbf{Expected market mechanism and interpretation} &
\textbf{Key sources} \\
\hline
\endfirsthead

\hline
\textbf{Period} &
\textbf{Episode and context} &
\textbf{Scale} &
\textbf{Expected market mechanism and interpretation} &
\textbf{Key sources} \\
\hline
\endhead

\hline
\multicolumn{5}{r}{\textit{Continued on next page}}\\
\endfoot

\hline
\endlastfoot

2007--2008 &
Global food price shock and price unrest in an import-dependent economy. &
Global \& Haiti &
External price transmission raises levels and increases dispersion as pass-through varies with import timing and distribution constraints. &
\citep{headey2010} \\

2010 (Jan.) &
Catastrophic earthquake causing large-scale infrastructure damage and long-lived disruption. &
National (urban core) &
A structural shock to logistics and market functioning increases intermittency in supply and produces persistent abnormal dispersion. &
\citep{usgs2010_haiti_eq} \\

2013--2017 &
Recurrent climate and hurricane impacts affecting livelihoods and access. &
National &
Disaster impacts disrupt trade flows and infrastructure and can generate sharp volatility spikes as corridor access fluctuates. &
\citep{ocha2017_haiti_irma} \\

2015--2016 &
Political disruption and electoral uncertainty, including delayed or disputed electoral processes. &
National &
Institutional instability and uncertainty raise dispersion through expectations shocks, episodic market closures, and interruptions to commerce. &
\citep{un_press2016_haiti_election_postponed} \\

2018 &
Fuel price announcement and riots with major disruption to mobility and commerce. &
National (urban) &
Transport-cost shocks and market interruptions generate discrete repricing and elevated dispersion. &
\citep{mcculloch2022} \\

2019 &
PetroCaribe protests and \emph{peyi l\`ok} lockdown disruptions. &
National &
Sustained disruption fragments markets and induces intermittent availability and abrupt corrections, generating prolonged elevated volatility. &
\citep{sanchezmartin2020} \\

2021 &
Presidential assassination and renewed political instability. &
National &
Political shock amplifies uncertainty and weakens market coordination, contributing to repeated interruptions and fragile expectations. &
\citep{ajil2023_moise_assassination} \\

2022 &
Fuel access crisis around Terminal Varreux and worsening access constraints. &
National (Port-au-Prince focus) &
Control of key infrastructure creates intermittent access to fuel and goods, driving volatility via distribution breakdown and localized shortages. &
\citep{ocha2022_haiti_fuel_access} \\

2023--2024 &
Escalating insecurity and corridor control; major violence surge and emergency conditions. &
National (urban core) &
Sustained market fragmentation generates persistent abnormal dispersion as supplies become sporadic and trading becomes discontinuous. &
\citep{icg2024} \\

\end{longtable}
\endgroup

Table~\ref{tab:haiti_mechanisms} summarizes the dominant interpretation channels associated with each event class. The goal is to standardize narrative interpretation across cases while keeping the framework descriptive rather than causal.

\begin{table}[htbp]
\centering
\begingroup
\footnotesize
\setlength{\tabcolsep}{3pt}
\renewcommand{\arraystretch}{1.07}
\sloppy

\begin{tabularx}{\textwidth}{@{}P{3.0cm} Y P{3.1cm}@{}}
\hline
\textbf{Event class} & \textbf{Typical volatility mechanism in market price data} & \textbf{Examples in Haiti} \\
\hline
Global price shock &
External transmission produces uneven pass-through and abrupt repricing in import-dependent urban markets. &
2007--08 crisis \\
\hline
Major disaster / structural damage &
Infrastructure collapse and logistics disruption create persistent intermittency and prolonged abnormal dispersion. &
2010 earthquake \\
\hline
Political disruption / unrest &
Uncertainty, market closures, and mobility constraints generate discontinuous trading and volatility clusters. &
2015--2016 disruptions; 2019 \emph{peyi l\`ok} \\
\hline
Fuel and logistics shock &
Transport-cost shocks transmit quickly through distribution margins and availability, producing sharp spikes. &
2018 riots; 2022 fuel access crisis \\
\hline
Sustained insecurity / corridor control &
Market fragmentation becomes structural, producing persistent high-volatility regimes rather than isolated spikes. &
2023--2024 violence surge \\
\hline
\end{tabularx}

\caption{Interpretation guidance for relating contextual events to detected volatility segments in Port-au-Prince.}
\label{tab:haiti_mechanisms}
\endgroup
\end{table}

\section{Supplementary Materials: Philippines Case Study (Sulu)}\label{app:philippines_events}

This appendix provides an annotated timeline of major episodes used to interpret volatility dynamics in selected Philippine markets. The timeline is compiled from disaster and humanitarian reporting, policy documentation, macro and cost transmission sources, and public records. These materials are used solely for \textit{ex post} interpretation of detected volatility segments and are not employed for causal identification.

\begingroup
\footnotesize
\setlength{\tabcolsep}{3pt}
\renewcommand{\arraystretch}{1.07}
\setlength{\LTleft}{0pt}
\setlength{\LTright}{0pt}
\sloppy

\begin{longtable}{@{}P{1.55cm} P{4.10cm} P{1.55cm} P{6.05cm} S{2.05cm}@{}}
\caption{Philippines (selected markets): annotated event timeline for interpreting detected volatility episodes.}\label{tab:philippines_timeline_full}\\
\hline
\textbf{Period} &
\textbf{Episode and context} &
\textbf{Scale} &
\textbf{Expected market mechanism and interpretation} &
\textbf{Key sources} \\
\hline
\endfirsthead

\hline
\textbf{Period} &
\textbf{Episode and context} &
\textbf{Scale} &
\textbf{Expected market mechanism and interpretation} &
\textbf{Key sources} \\
\hline
\endhead

\hline
\multicolumn{5}{r}{\textit{Continued on next page}}\\
\endfoot

\hline
\endlastfoot

2007--2008 &
Global food and fuel price crisis transmitted to domestic staple markets. &
Global \& Philippines &
External price transmission raises levels and increases dispersion via uneven pass-through, procurement timing, and heterogeneous substitution and distribution margins. &
\citep{dawe2010} \\

2012 (Oct.) &
Framework Agreement on the Bangsamoro and heightened political mobilization in Moro areas, including the Sulu archipelago. &
Subnational \& national &
Political uncertainty and episodic security risks raise corridor premia and reduce market integration, increasing dispersion through intermittent inflows and discontinuous trading. &
\citep{icg2012_breakthrough_mindanao} \\

2012 (Dec.) &
Typhoon Bopha (Pablo) strikes Mindanao, causing widespread destruction and disruption to production and transport. &
Subnational (Mindanao) &
A disaster shock disrupts supply chains and access, creating localized shortages and abrupt price jumps with partial corrections as flows resume. &
\citep{ocha2012_pablo_sitrep2} \\

2013 &
Zamboanga crisis and Typhoon Haiyan with large displacement and logistics impacts. &
National \& subnational &
Disasters and localized conflict disrupt supply chains, producing abrupt dislocations and volatility spikes as corridor access fluctuates and distribution resumes. &
\citep{ocha2013_haiyan} \\

2017 &
Marawi siege with displacement and movement restrictions and broader trade spillovers. &
Subnational (Mindanao) &
Security disruptions fragment market access and raise dispersion through intermittent inflows, mobility constraints, and nonlinear adjustment of distribution margins. &
\citep{ocha2017_marawi} \\

2014--2018 &
Energy and freight-cost volatility with renewed price pressure during the 2018 fuel upswing. &
Global \& Philippines &
Energy and freight costs transmit through distribution margins; shifts in transport costs can generate step changes and elevated dispersion. &
\citep{abs2018} \\

2019 (Feb.) &
Rice Tariffication Law (RA~11203) and discontinuity in the rice import regime. &
National (policy) &
Policy shifts alter price expectations and market structure, generating adjustment dynamics and dispersion as supply conditions re-optimize. &
\citep{officialgazette2019_ra11203} \\

2020 (Jan.) &
Taal volcano eruption disrupting activity and logistics. &
National &
Disaster shock temporarily disrupts flows and access, raising dispersion as supply conditions adjust unevenly. &
\citep{ndrrmc2020_taal} \\

2020 (Mar.) &
COVID-19 pandemic and associated mobility and supply-chain disruption. &
National &
Movement constraints and uncertainty disrupt logistics and distribution margins, generating volatility spikes through episodic market adjustment. &
\citep{worldbank2021_phl} \\

2022 &
Global energy and shipping-cost pressures following the Russia--Ukraine shock. &
Global \& Philippines &
Externally driven cost shocks transmit through fuel and shipping, raising dispersion via uneven pass-through and time-varying supply conditions. &
\citep{fao2022_ukraine} \\

2023 &
Persistently elevated inflation without a single discrete trigger. &
National &
Sustained cost pressures can produce clusters of high-volatility flags driven by repeated repricing and heterogeneous adjustment across markets. &
\citep{imf2023_philippines_articleiv} \\

\end{longtable}
\endgroup

Table~\ref{tab:philippines_mechanisms} summarizes the dominant interpretation channels associated with each event class. The goal is to standardize narrative interpretation across cases while keeping the framework descriptive rather than causal.

\begin{table}[htbp]
\centering
\begingroup
\footnotesize
\setlength{\tabcolsep}{3pt}
\renewcommand{\arraystretch}{1.07}
\sloppy

\begin{tabularx}{\textwidth}{@{}P{3.0cm} Y P{3.1cm}@{}}
\hline
\textbf{Event class} & \textbf{Typical volatility mechanism in market price data} & \textbf{Examples in the Philippines} \\
\hline
Global price / cost transmission &
Uneven pass-through of international food, fuel, and shipping costs raises dispersion via procurement timing and heterogeneous distribution margins. &
2007--08; 2022 \\
\hline
Disaster / logistics disruption &
Infrastructure damage and mobility constraints generate discontinuous trading and sharp volatility spikes. &
Haiyan (2013); Taal (2020) \\
\hline
Localized security disruption &
Movement restrictions and access constraints fragment markets and produce intermittent inflows. &
Marawi siege (2017) \\
\hline
Policy discontinuity &
Regime changes alter expectations and supply behavior, producing adjustment dynamics and elevated dispersion. &
Rice Tariffication Law (2019) \\
\hline
Persistent cost pressure &
Sustained inflation can produce clustered high-volatility flags even without a single focal trigger. &
2023 inflation persistence \\
\hline
\end{tabularx}

\caption{Interpretation guidance for relating contextual events to detected volatility segments in selected Philippine markets.}
\label{tab:philippines_mechanisms}
\endgroup
\end{table}

\end{document}